\documentclass[epj]{svjour}
\usepackage{graphics}
\usepackage[utf8]{inputenc}
\usepackage{graphicx}
\usepackage{psfrag}
\usepackage{amssymb}
\usepackage{amsmath}
\usepackage{epstopdf}
\usepackage{color}
\usepackage{bm}
\usepackage{dcolumn}
\usepackage{xcolor}
\usepackage{url}
\usepackage{subfigure}
\usepackage{multirow}
\begin{document}
\title{An Analysis of Parton Distribution Functions of the Pion and the Kaon with the Maximum Entropy Input}
\author{Chengdong Han\inst{1,2}
\and Gang Xie\inst{1,3}
\and Rong Wang\inst{1,2}\thanks{\emph{Email address:} rwang@impcas.ac.cn}
\and Xurong Chen\inst{1,2,3}\thanks{\emph{Email address:} xchen@impcas.ac.cn}
%
}                     
\offprints{}          
\institute{Institute of Modern Physics, Chinese Academy of Sciences, Lanzhou 730000, China
\and School of Nuclear Science and Technology, University of Chinese Academy of Sciences, Beijing 100049, China
\and Guangdong Provincial Key Laboratory of Nuclear Science, Institute of Quantum Matter, South China Normal University, Guangzhou 510006, China }
\date{Received: date / Revised version: date}
%
\abstract{
We present pion and kaon parton distribution functions from a global QCD analysis
of the experimental data within the framework of dynamical parton model.
We use the DGLAP equations with parton-parton recombination corrections
and the valence input of uniform distribution which maximizes the information entropy.
At our input scale $Q_0^2$, there are no sea quark and gluon distributions.
All the sea quarks and gluons of the pion and the kaon are completely generated from the parton splitting processes.
The mass-dependent parton splitting kernel is applied for the strange quark distribution in the kaon.
The obtained valence quark and sea quark distributions at high $Q^{2}$ ($Q^2>5$ GeV$^2$)
are compatible with the existed experimental measurements.
Furthermore, the asymptotic behaviours of parton distribution functions
at small and large $x$ have been studied for both the pion and the kaon.
Lastly, the first three moments of parton distributions at high $Q^{2}$ scale are calculated,
which are consistent with other theoretical predictions.
} 
\maketitle

\section{Introduction}
\label{sec:intro}

Pion and kaon, as the pseudo-Goldstone bosons \cite{Nambu:1960tm,Goldstone:1962es}, are the important objects
for understanding the dynamical chiral symmetry breaking.
As the mass low-lying mesons, their internal structures are crucial for us
to see the emergence of hadron mass from quantum chromodynamics (QCD)
of strong interaction \cite{Maris:1997hd,Roberts:2019ngp,Roberts:2020udq,Chen:2020ijn},
together with the inquiring of the origin of proton mass \cite{Ji:1994av,Yang:2018nqn,Lorce:2017xzd,Chen:2020gml}.
Pion, as the lightest meson, plays a dominant role
in mediating the nuclear force. Kaon, having a heavier strange quark inside,
is interesting for us to see the interference between the emergence of mass from QCD
and the Higgs mechanism \cite{Cui:2020dlm}. The study on pion and kaon structures
is a way to test the nonperturbative QCD approaches,
and to help answering the fundamental question on the origin of hadron mass.
Parton distribution functions (PDFs) display the longitudinal momentum structure
of the quarks and gluons in the infinite momentum frame.
In theory, there are quite a lot of methods giving pion
and kaon PDFs, such as lattice QCD (LQCD) \cite{Detmold:2003tm,Shugert:2020tgq,Sufian:2020vzb,Joo:2019bzr,Gao:2020ito},
Dyson-Schwinger Equations (DSE) \cite{Nguyen:2011jy,Chen:2016sno,Ding:2019lwe},
holographic QCD \cite{deTeramond:2018ecg}, light-front quantization \cite{Lan:2019vui},
and the constituent quark model \cite{Szczepaniak:1993uq,Frederico:1994dx,Watanabe:2016lto}.

In experiment, much less is known for the internal structures
of the pion and the kaon, due to the absence of the fixed pion and kaon targets.
The current available data come from the pion- or kaon- induced Drell-Yan (DY) reaction
on the nuclear target at CERN (NA3 and NA10) \cite{Badier:1980jq,Badier:1983mj,Betev:1985pf} and Fermilab (E615) \cite{Conway:1989fs},
and the leading-neutron deep inelastic scattering (LN-DIS) of $e$-$p$ collision at HERA (ZEUS and H1) \cite{Chekanov:2002pf,Aaron:2010ab}.
Thanks to the ZEUS and H1 measurements at small $x$,
the sea quark and gluon distributions are much better constrained for the pion.
For the kaon, the data is very scarce and they are from the NA3 DY measurement more than 40 years ago.
More precise measurements from the DY process in the future are longed.
The small-$x$ data of the pion and the kaon with a better kinematical coverage to the current
DY data from LN-DIS are also highly needed.

Although the structure data on pion and kaon are scarce,
there have been some pioneering works on the global QCD analyses of the pion PDFs.
The previous analyses of GRS \cite{Gluck:1999xe} and SMRS \cite{Sutton:1991ay}
constrain well the pion PDFs in the valence region using
the dilepton (DY) and the prompt photon production data in $\pi$-nucleus scattering.
Recently, JAM collaboration \cite{Barry:2018ort} includes both the DY data
and the LN-DIS data in the analysis.
By using a Bayesian Monte Carlo method, JAM collaboration finds that
by including the LN-DIS data from HERA,
the sea quark and gluon distributions are better constrained.
The gluon distribution at small $x$ is found to be much different from the analysis
inferred from the DY data alone. The other significant finding is that the pion PDFs
from the global analysis describe amazingly well the $\bar{d}-\bar{u}$ asymmetry in the proton \cite{Towell:2001nh},
using the same pion cloud model for the LN-DIS analysis \cite{Barry:2018ort}.
Recently xFitter collaboration \cite{Novikov:2020snp} finds that the current data are not enough
to determine sea quark and gluon distributions unambiguously.
Moreover, they explored the theoretical uncertainties from such as varying the renormalization and factorization scales,
and from the flexibility of the chosen parametrization.
The conclusion is that the valence distribution is well constrained, while the total uncertainty
of experimental and theoretical uncertainties are large for the obtained sea quark and gluon distributions.
Last but not least, another interesting analysis of the DY data only finds that the power law of $(1-x)^{\beta}$
at $x\rightarrow 1$ limit agrees with the perturbative QCD prediction \cite{Ezawa:1974wm,Farrar:1975yb,Berger:1979du,Chang:2020kjj}
when the soft-gluon resummation correction is considered \cite{Aicher:2010cb}.

In this work, to better fix the sea quark and gluon distributions,
we perform a global analysis under the framework
of dynamical parton model \cite{Gluck:1977ah,Wang:2016sfq}.
In our dynamical parton model, the sea quark and gluon distributions are purely generated
in QCD evolution, which means that there are no intrinsic sea quarks and gluons
at the sufficiently low scale.
Starting from a very low scale $Q_0^2\sim 0.1$ GeV$^2$,
we use the state-of-the-art process-independent renormalization-group-invariant
strong coupling $\alpha_{\rm s}$ in the analysis \cite{Cui:2019dwv}.
This type of strong coupling saturates in the infrared region.
In the low resolution scale, the parton overlapping and the parton-parton recombination effect
cannot be ignored \cite{Wang:2016sfq,Gribov:1984tu,Mueller:1985wy,Chen:2013nga}.
Therefore, we use the parton-parton recombination corrected DGLAP equations
to slow down the parton splitting processes in low $Q^2$ region.
For QCD evolution in terms of strange quark distribution,
the mass-dependent DGLAP kernel \cite{Cui:2020dlm,Chang:2014lva}
is taken into account since the strange quark mass cannot be ignored at low $Q^2$.
We present also the difference between the strange valence distribution
and the up valence distribution in the kaon
caused by the mass correction to the parton splitting kernel \cite{Cui:2020dlm,Chang:2014lva}.
Finally, we investigate the large-$x$ and small-$x$ behaviors of pion and kaon PDFs.

The organization of the paper is as follows.
Section \ref{sec:MEM-input} introduces the nonperturbative input used in this global analysis
based on the dynamical parton model.
In Sec. \ref{sec:evolution-Eq}, we discuss the DGLAP equations with parton-parton recombination corrections
and the saturating running strong coupling $\alpha_{\rm s}$.
Section \ref{sec:global-fit} shows the result of the least-square fit,
the optimal hadronic scale $Q_0^2$,
and the optimal $R$ for the magnitude of two-parton recombination process.
Section \ref{sec:result} illustrates the global fitting results compared with
the experimental data, the resulting large-$x$ and small-$x$ behaviors of the PDFs of the pion and the kaon,
and the obtained dynamical sea quark and gluon distributions.
At the end, a short summary is given in Sec. \ref{sec:summary}.

\section{Nonperturbative input from maximum entropy method}
\label{sec:MEM-input}

In the global analysis, the nonperturbative input is the assumed
initial PDFs at $Q_0^2$. The PDFs at any higher
$Q^2$ are computed from the nonperturbative input with QCD evolution
equations which describe the $Q^2$-dependence of PDFs.
The first task for the global analysis is to find the optimal nonperturbative
input which reproduces well the experimental data measured at the hard scale ($Q^2\gtrsim 5$ GeV$^2$).
In the dynamical parton model, the input scale $Q_0^2$ is usually chosen to be
closer to the hadronic scale ($\sim 0.1$ GeV$^2$), where only the valence quarks
(constituent quarks) can be resolved in the hadron. There are two benefits of using
the dynamical parton model. First, the nonperturbative input is related
to the simple quark model predictions. Second, the sea quark and gluon distributions
are completely produced from QCD evolution equations,
which indicates the dynamical origin of the gluon and sea quarks.

At the hadronic scale, the pion plus has only one up quark and one anti-down quark.
Therefore we have the following valence number sum rule and momentum sum rule for $\pi^{+}$:
\begin{equation}
\begin{split}
\int_0^1 u_{\rm V}^{\rm \pi}(x,Q_0^2) dx = 1,\\
\int_0^1 \bar{d}_{\rm V}^{\rm \pi}(x,Q_0^2) dx = 1,\\
\int_0^1 x [ u_{\rm V}^{\rm \pi}(x,Q_0^2) + \bar{d}_{\rm V}^{\rm \pi}(x,Q_0^2) ] dx = 1.\\
\end{split}
\label{eq:pion-input-constraints}
\end{equation}
Similarly we have the following constraints for the nonperturbative input of $K^{+}$ at the hadronic scale:
\begin{equation}
\begin{split}
\int_0^1 u_{\rm V}^{\rm K}(x,Q_0^2) dx = 1,\\
\int_0^1 \bar{s}_{\rm V}^{\rm K}(x,Q_0^2) dx = 1,\\
\int_0^1 x [ u_{\rm V}^{\rm K}(x,Q_0^2) + \bar{s}_{\rm V}^{\rm K}(x,Q_0^2) ] dx = 1.\\
\end{split}
\label{eq:kaon-input-constraints}
\end{equation}
The above nonperturbative inputs have the minimum number of components inside the hadrons.
We need to compose the functions for the nonperturbative input.

Naively the initial parton distributions at $Q_0^2$
can be estimated with the statistical models \cite{Bourrely:2018yck,Bourrely:2020izp,Wang:2014lua,Han:2018wsw}.
The quantum statistical approach gives the pion parton distributions
in agreement with the DY data, using Fermi-Dirac parametric form for quark distribution
and allowing some parameters for
the thermodynamical potential and temperature \cite{Bourrely:2018yck,Bourrely:2020izp}.
Our previous works show that the maximum entropy method (MEM)
gives the reasonable input valence quark distributions
for the proton and the pion \cite{Wang:2014lua,Han:2018wsw}.
Applying MEM and taking the charge symmetry $u_{\rm V}^{\rm \pi}(x,Q_0^2) = \bar{d}_{\rm V}^{\rm \pi}(x,Q_0^2)$,
the nonperturbative input is evaluated to be the uniform distribution \cite{Wang:2014lua,Han:2018wsw}, which is written as,
\begin{equation}
\begin{split}
u_{\rm V}^{\rm \pi}(x,Q_0^2) = \bar{d}_{\rm V}^{\rm \pi}(x,Q_0^2) = 1.
\end{split}
\label{eq:pion-input}
\end{equation}
In the maximum entropy method, the information entropy of the initial quark distributions is
taken to be at the maximum. Assuming $SU(3)$ flavor symmetry,
the nonperturbative input of the kaon is also the uniform distribution from the maximum entropy approach,
which is written as,
\begin{equation}
\begin{split}
u_{\rm V}^{\rm K}(x,Q_0^2) = \bar{s}_{\rm V}^{\rm K}(x,Q_0^2) = 1.
\end{split}
\label{eq:kaon-input}
\end{equation}

With the nonperturbative input of MEM, the following task is
to determine the input hadronic scale $Q_0^2$ for the input distribution.
For the proton, the hadronic scale with only valence components is around $Q_0^2=0.07$ GeV$^2$ \cite{Wang:2016sfq},
from a global analysis of the $l$-$p$ DIS data.
The hadrons of different sizes may have different initial resolution scales $Q_0^2$.
Therefore in this analysis of PDFs of pseudoscalar mesons,
the input hadronic scale is treated as a free parameter which should be fixed from the global fit.

\section{DGLAP equations with parton-parton recombination corrections}
\label{sec:evolution-Eq}

To connect the nonperturbative input with the experimental measurements
in the asymptotic region, the bridge is QCD evolution equations.
Usually, the Dokshitzer-Gribov-Lipatov-Altarelli-Parisi (DGLAP)
equations \cite{Dokshitzer:1977sg,Gribov:1972ri,Altarelli:1977zs}
are used to describe the PDF evolution over $Q^2$.
For the parton evolution starting from a very low $Q_0^2$,
the parton-parton recombination corrections should not be neglected.
The parton-parton recombination effect was first put forward by Gribov, Levin and Ryskin (GLR) \cite{Gribov:1984tu},
then followed by Mueller, Qiu (MQ) evaluating the recombination coefficients
with the covariant field theory \cite{Mueller:1985wy}.
The full recombination corrections (gluon-gluon, quark-gluon, quark-quark)
were derived by Zhu, Ruan, and Shen using the time ordered field theory \cite{Zhu:1998hg,Zhu:1999ht,Zhu:2004xj}.

The DGLAP equations with parton-parton recombination corrections used in this work
are given by \cite{Wang:2016sfq,Chen:2013nga},
\begin{equation}
\begin{aligned}
Q^2\frac{dxf_{q_i}(x,Q^2)}{dQ^2}
=\frac{\alpha_s(Q^2)}{2\pi}P_{qq}\otimes f_{q_i},
\end{aligned}
\label{eq:DGLAP-NS}
\end{equation}
for valence quark distributions,
\begin{equation}
\begin{aligned}
Q^2\frac{dxf_{\bar{q}_i}(x,Q^2)}{dQ^2}
=\frac{\alpha_s(Q^2)}{2\pi}[P_{qq}\otimes f_{\bar{q}_i}+P_{qg}\otimes f_g]\\
-\frac{\alpha_s^2(Q^2)}{4\pi R^2Q^2}\int_x^{1/2} \frac{dy}{y}xP_{gg\to \bar{q}}(x,y)[yf_g(y,Q^2)]^2\\
+\frac{\alpha_s^2(Q^2)}{4\pi R^2Q^2}\int_{x/2}^{x}\frac{dy}{y}xP_{gg\to \bar{q}}(x,y)[yf_g(y,Q^2)]^2,
\end{aligned}
\label{eq:DGLAP-S}
\end{equation}
for sea quark distributions, and
\begin{equation}
\begin{aligned}
Q^2\frac{dxf_{g}(x,Q^2)}{dQ^2}
=\frac{\alpha_s(Q^2)}{2\pi}[P_{gq}\otimes \Sigma+P_{gg}\otimes f_g]\\
-\frac{\alpha_s^2(Q^2)}{4\pi R^2Q^2}\int_x^{1/2} \frac{dy}{y}xP_{gg\to g}(x,y)[yf_g(y,Q^2)]^2\\
+\frac{\alpha_s^2(Q^2)}{4\pi R^2Q^2}\int_{x/2}^{x}\frac{dy}{y}xP_{gg\to g}(x,y)[yf_g(y,Q^2)]^2,
\end{aligned}
\label{eq:DGLAP-G}
\end{equation}
for gluon distribution,
where $P_{qq}$, $P_{qg}$, $P_{gq}$, $P_{gg}$ are the standard parton splitting
kernels, and $P_{gg\to \bar{q}}$, $P_{gg\to g}$ are the gluon-gluon
recombination coefficients which can be found in the literatures \cite{Zhu:1998hg,Zhu:1999ht,Zhu:2004xj}.
The $1/(4\pi R^2)$ in the evolution equations is for the two-parton density normalization,
and $R$ is viewed as the correlation length of two interacting partons.
In a previous analysis of the proton \cite{Wang:2016sfq}, the two-parton correlation length $R$
is fitted to be 3.98 GeV$^{-1}$.
For different hadrons, the values of $R$ are different and related to the radii of the hadrons.
In this analysis of the pion and the kaon, $R$ is a free parameter to be
determined with the global fit to the data.

\begin{figure}[htbp]
\centering
\includegraphics[scale=0.39]{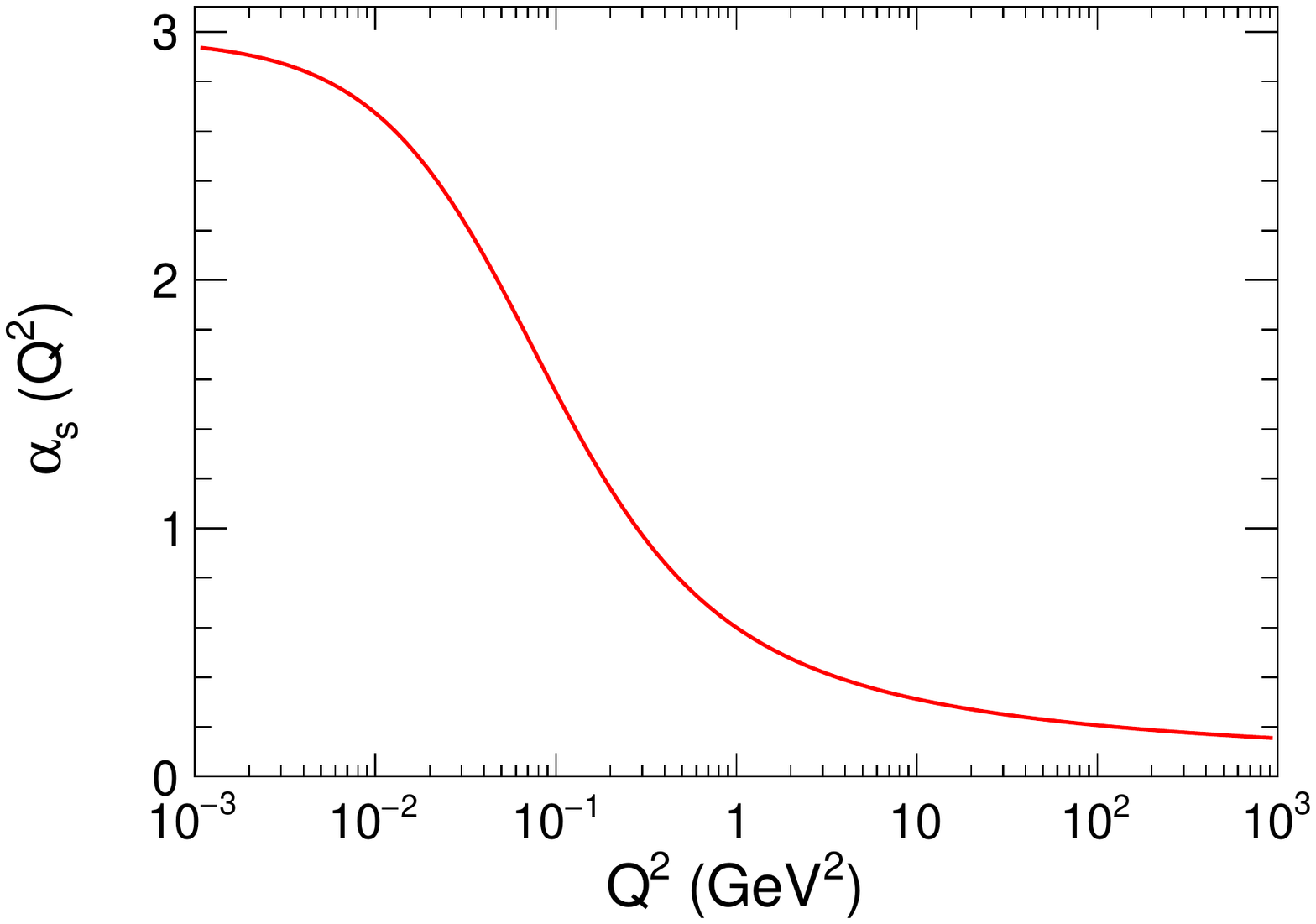}
\caption{The infrared-saturating strong coupling $\alpha_{\rm s}$ \cite{Cui:2019dwv} used in this work. }
\label{fig:storng-coupling}
\end{figure}

The running strong coupling constant $\alpha_{s}$ is an important parameter
in QCD evolution equations. In this work, we use a renormalization-group-invariant
process-independent effective strong coupling constrained by the calculation of LQCD \cite{Cui:2019dwv}.
The effective strong coupling shows a saturated plateau approaching the infrared region,
which agrees well with the experimental measurement of Bjorken sum rule at low $Q^2$ ($<1$ GeV$^2$).
The saturated effective strong coupling is given by \cite{Cui:2019dwv},
\begin{equation}
\begin{split}
\alpha_{\rm s}(Q^{2}) = \frac{4\pi}{\beta_{0}{\rm Ln}[(m_{\alpha}^{2} + Q^{2}) / \Lambda^{2}_{\rm QCD}]},
\end{split}
\label{eq:strong-coupling}
\end{equation}
where $\beta_{0}=(33-2n_{\rm f})/3$ refers to the one-loop $\beta$ function coefficient,
$n_{\rm f}$ is the number of flavors,
$m_{\alpha}=0.43$ GeV is the effective gluon mass owning to the dynamical breaking of scale invariance \cite{Cui:2019dwv}.
For the QCD cutoff, we chose $\Lambda_{\rm QCD}=0.34$ GeV.
Fig. \ref{fig:storng-coupling} shows the running strong coupling applied in this analysis,
which has the strong coupling $\alpha_s$ saturated around 3.

\section{Determinations of the optimal parameters and the method of PDF error analysis}
\label{sec:global-fit}

As the nonperturbative input is taken as the MEM input,
only the hadronic scale $Q_0^2$ and the parton-parton recombination parameter $R$ are
left to be determined by the global analysis.
We use the least-square fit to find the optimal parameters
for the pion and the kaon respectively. The reduced $\chi^2$ for the fit is defined as,
\begin{equation}
\begin{split}
\chi^2 / {\rm ndf} = \frac{1}{N_{\rm tot.}-n_{\rm par.}}\sum^{N_{\rm exp.}}_{i}
\sum_{j}^{N_{i}}\frac{(D_{j}-T_{j})^{2}}{\sigma_{j}^{2}},
\end{split}
\label{eq:chi2-def}
\end{equation}
in which $N_{\rm tot.}$ is the total number of data points of all experiments,
$n_{\rm par.}=2$ is the number of free parameters in our analysis.
$N_{\rm exp.}$ is the number of experimental measurements used in the analysis;
$N_{i}$ is the number of data points for experiment i;
$D_{j}$ is one data point for the experimental observables;
$\sigma_{j}$ is the corresponding error of the data point $D_{j}$;
And $T_{j}$ is the corresponding theoretical prediction of $D_{j}$.
Specifically for this study, the experimental observables $D_{j}$ are the valence quark distribution
measured in DY reaction, and the structure function $F_2$ measured in  LN-DIS process.
For the least-square fit, we use the MINUIT package to find the minimum $\chi^2$
and the optimal parameters.

To analyze the statistical uncertainties of PDFs and the related quantities,
we chose the Hessian method \cite{Pumplin:2001ct,Martin:2002aw},
which is a convenient and efficient method
for calculating the uncertainty due to the errors of the experimental data.
The Hessian approach is based on the quadratic approximation of the $\chi^2$
function in the neighborhood of the minimum (the best fit).
We can write,
\begin{equation}
\begin{split}
\Delta\chi^2=\chi^2-\chi_0^2=\sum_{i=1}^{n}\sum_{j=1}^{n}H_{ij}(a_i-a_i^0)(a_j-a_j^0),
\end{split}
\label{eq:HessianMatrix}
\end{equation}
where $H_{ij}$ is the element of the Hessian matrix,
$a_i$ are the model parameters specifying the PDFs,
$a_i^0$ are the fitted values of the parameters at minimum $\chi^2$,
and $n$ is the number of parameters.
The uncertainty of any quantity $X$ can be calculated using the standard
linear propagation of errors, which is written as,
\begin{equation}
\begin{split}
(\Delta X)^2 = \Delta\chi^2 \sum_{i=1}^{n}\sum_{j=1}^{n} \frac{\partial X}{\partial a_i}C_{ij}(a)\frac{\partial X}{\partial a_j},
\end{split}
\label{eq:ErrorPropagation}
\end{equation}
where $C_{ij}(a)=(H^{-1})_{ij}$ is the error matrix of the parameters, and $\Delta\chi^2$
is the allowed variation of $\chi^2$.
In this work, we take $\Delta\chi^2=11.8$ for 3$\sigma$ error with two free parameters.
Usually it is more convenient to calculate the error
using the diagonalized Hessian matrix and the eigenvectors.

In the analysis of pion PDFs, we take the DY data from Fermilab E615 Collaboration \cite{Conway:1989fs}
and the structure function data from H1 Collaboration at HERA \cite{Aaron:2010ab}. For the analysis of kaon PDFs,
we take the $u^{\rm K}/u^{\pi}$ data from CERN-NA3 experiment \cite{Badier:1980jq}.
With the least-square fit, we get the hadronic scales of the nonperturbative inputs for the pion
and the kaon, and the two-parton correlation length $R$, which are listed in table \ref{tab:fit-results}.
The qualities of the two fits are amazingly good ($\chi^2$/ndf$<1$).
Only the experimental data at small $x$ ($<0.01$) are sensitive to
the parameter $R$ for the parton-parton recombination strength.
The CERN-NA3 data used to constrain kaon PDFs are at the large $x$ ($>0.2$).
Therefore it is impossible to determine the parameter $R$ for the kaon with accuracy.
The small-$x$ data on the kaon structure are needed,
which could be accessed with the future electron-ion colliders \cite{Chen:2020ijn,Chen:2018wyz,Accardi:2012qut}.
In this work, we let the parameter $R$ to be the same for both the pion and the kaon,
for they are in the same meson octet and both of the Glodstone boson nature.

In this analysis, the hadronic scale is determined to be $0.129\pm 0.003$ GeV$^{2}$ for the pion,
and the hadronic scale is determined to be $0.110\pm 0.006$ GeV$^{2}$ for the kaon.
In the previous and similar analysis of the proton using the pure valence quark input,
the hadronic scale is 0.067 GeV$^{2}$ \cite{Wang:2016sfq}.
The smaller the hadron size is, the higher resolution scale is required
to just resolve the minimum quark components of the hadron.
Since the sizes of pion and kaon are smaller than that of proton,
the input hadronic scales should be higher than that of proton.
The obtained hadronic scales for pion and kaon are within the naive expectations.
The parton-parton correlation length for pion and kaon is obtained to be $R=2.39\pm 0.08$ GeV$^{-1}$ in this work,
which is smaller than that of proton (3.98 GeV$^{-1}$) \cite{Wang:2016sfq}.
This also could be understood as resulting from the smaller radius of the meson,
since the correlation length $R$ is proportional to the radius of the hadron.
The obtained input hadronic scales $Q_0^2$ and parton-parton recombination parameter $R$
for the pion and the kaon are within the theoretical implications.

\begin{table}
  \caption{
    The obtained input scale $Q_{0}^{2}$ and the parton-parton correlation length $R$ for the pion
    and the kaon, from the fittings to the popular experimental data up to date.
    The errors in the table indicate the 1$\sigma$ uncertainties.
  }
  \label{tab:fit-results}
  \begin{tabular}{cccccc}
  \hline\hline
            & $Q^{2}_{0}$/GeV$^{2}$  & $R$/GeV$^{-1}$   & $\chi^2$/ndf & Experiment & $N$ points \\
  \hline
    $\pi$   & 0.129(3)                & 2.39(8)            & $0.864$    &  E615 + H1 & 40 + 29  \\
    K       & 0.110(6)                & 2.39(8)           & $0.750$    &  CERN-NA3  & 8        \\
  \hline\hline
  \end{tabular}
\end{table}

\section{Results and discussions}
\label{sec:result}

\begin{figure}[htbp]
\centering
\includegraphics[scale=0.41]{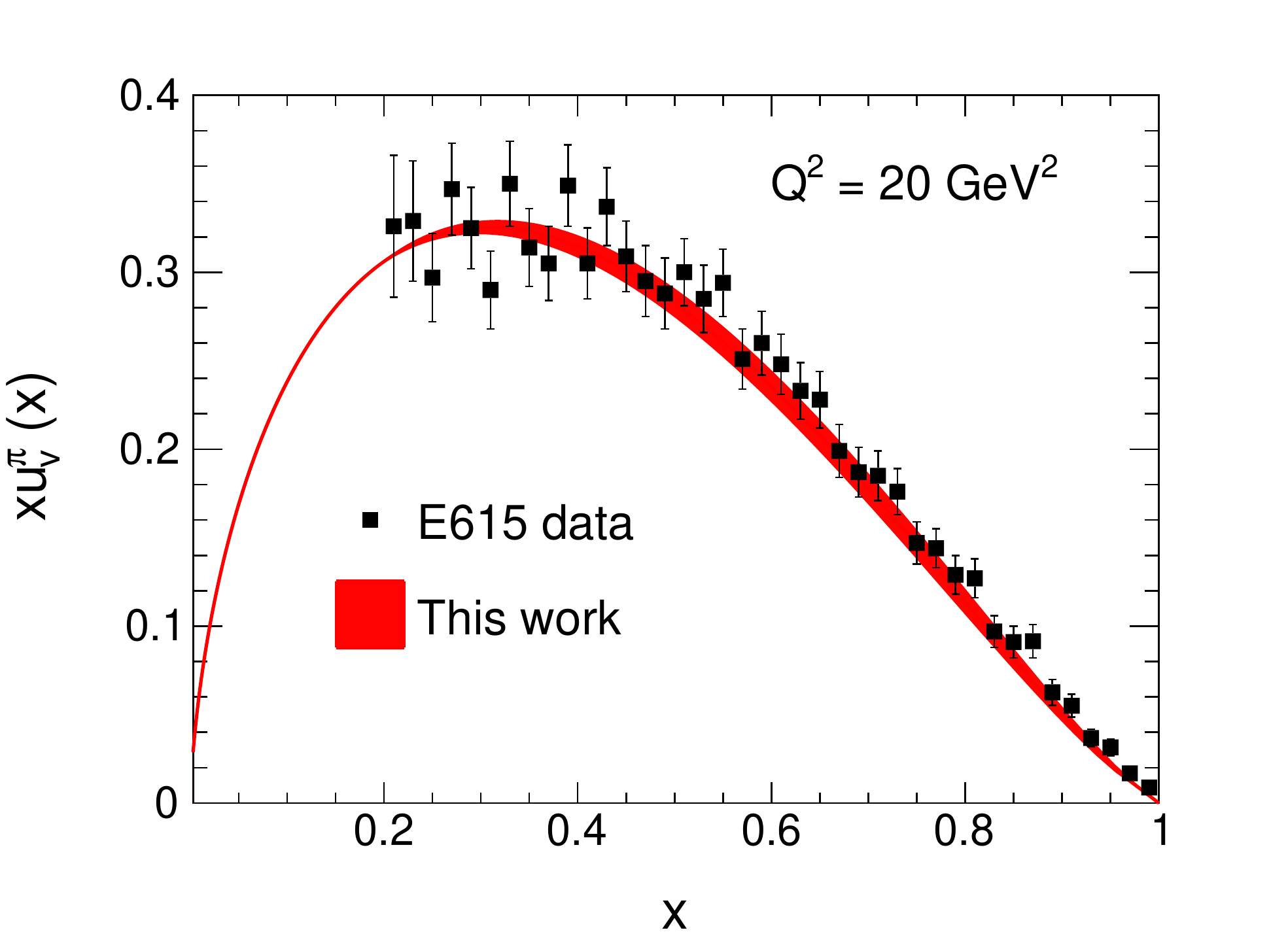}
\caption{Comparisons of our predicted up valence quark distribution of the pion with E615 experimental data \cite{Conway:1989fs}.
The error band of this work shows the 3$\sigma$ uncertainty of the valence distribution. }
\label{fig:xuv-E615}
\end{figure}

\begin{figure}[htbp]
\centering
\includegraphics[scale=0.41]{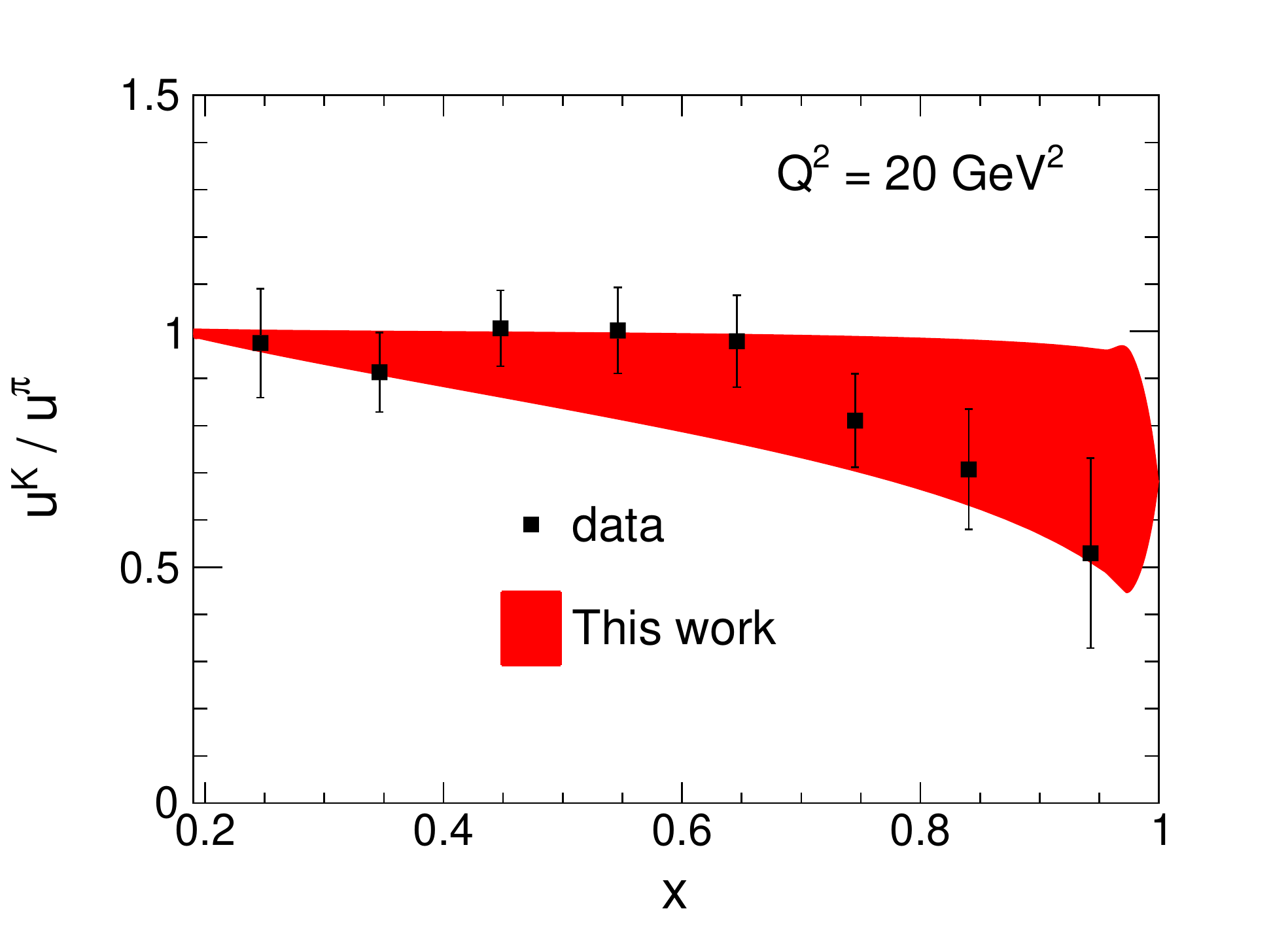}
\caption{Comparisons of our predicted ratio $u^{\rm K}/u^{\pi}$ as a function of $x$ with CERN-NA3 experimental data \cite{Badier:1980jq}.
The error band of this work shows the 3$\sigma$ uncertainty of the ratio. }
\label{fig:upi-over-uK-NA3}
\end{figure}

With the obtained parameters, we calculate pion and kaon PDFs with QCD evolution equations and their errors.
Fig. \ref{fig:xuv-E615} shows our obtained pionic valence quark distribution compared with the Drell-Yan data
from Fermilab E615 experiment \cite{Conway:1989fs}.
Fig. \ref{fig:upi-over-uK-NA3} shows our obtained ratio between kaon up quark distribution
and pion up quark distribution, compared with CERN-NA3 Drell-Yan data from the pion and the kaon beams \cite{Badier:1980jq}.
We find that both pion and kaon PDFs determined in this work are consistent with the experimental measurements
in the large-$x$ region ($x>0.2$).
Our prediction of $u^{\rm K}/u^{\pi}$ goes down smaller than one as $x$ goes from 0.2 to 1.
The CERN-NA3 experiment shows that the ratio $u^{\rm K}/u^{\pi}$ is smaller than one when $x>0.7$.
We expect more and precise experimental data to figure out where the kaon up quark distribution
starts to be smaller than the pion up quark distribution.

\begin{figure*}[htbp]
\centering
\includegraphics[scale=0.85]{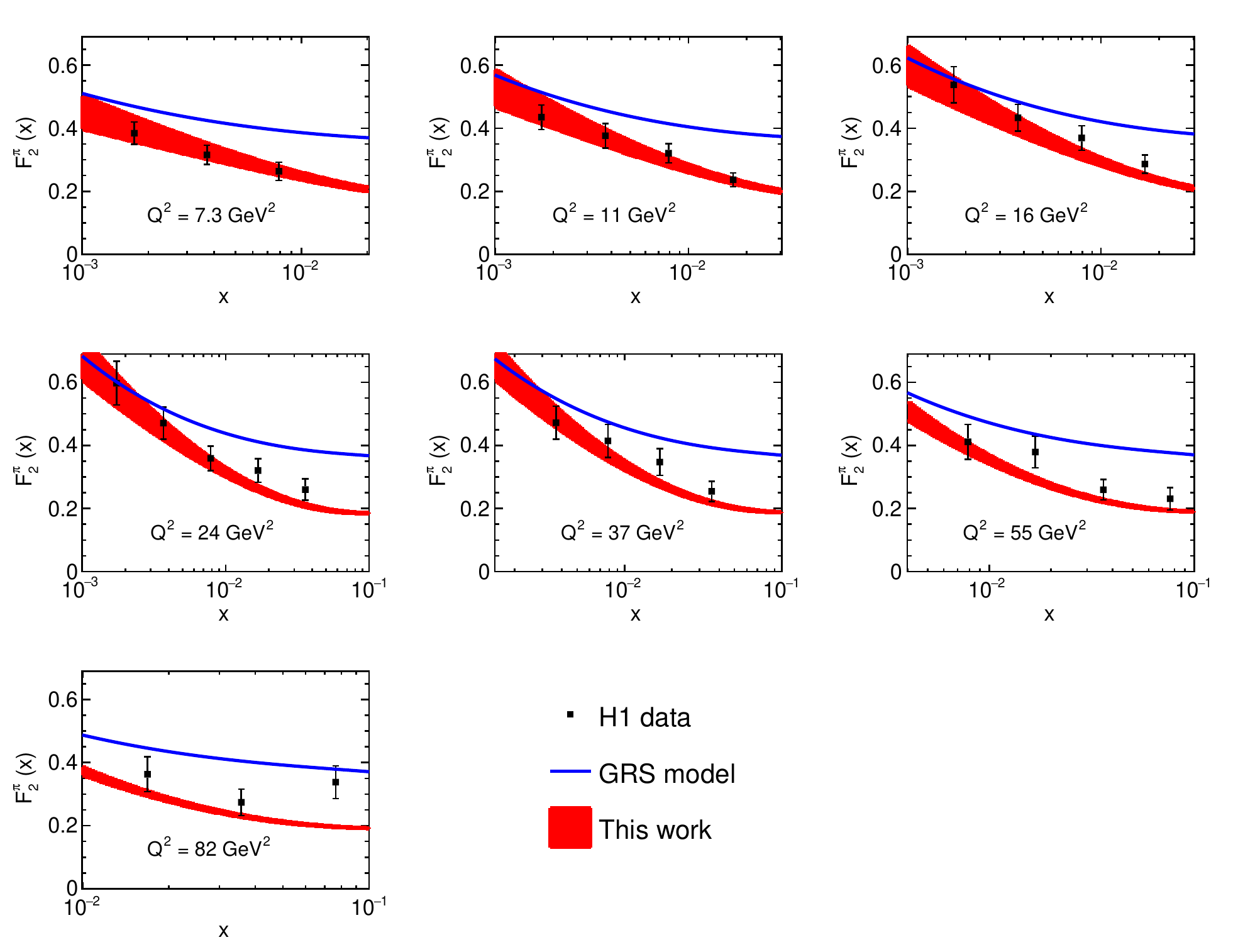}
\caption{Our predicted structure function $F_{2}^{\pi}(x, Q^{2})$ compared with H1 data \cite{Aaron:2010ab},
and GRS model predictions \cite{Gluck:1999xe}.
The error bands of this work show the 3$\sigma$ uncertainties of the structure function. }
\label{fig:F2pi-H1}
\end{figure*}

\begin{figure*}[htbp]
\centering
\includegraphics[scale=0.85]{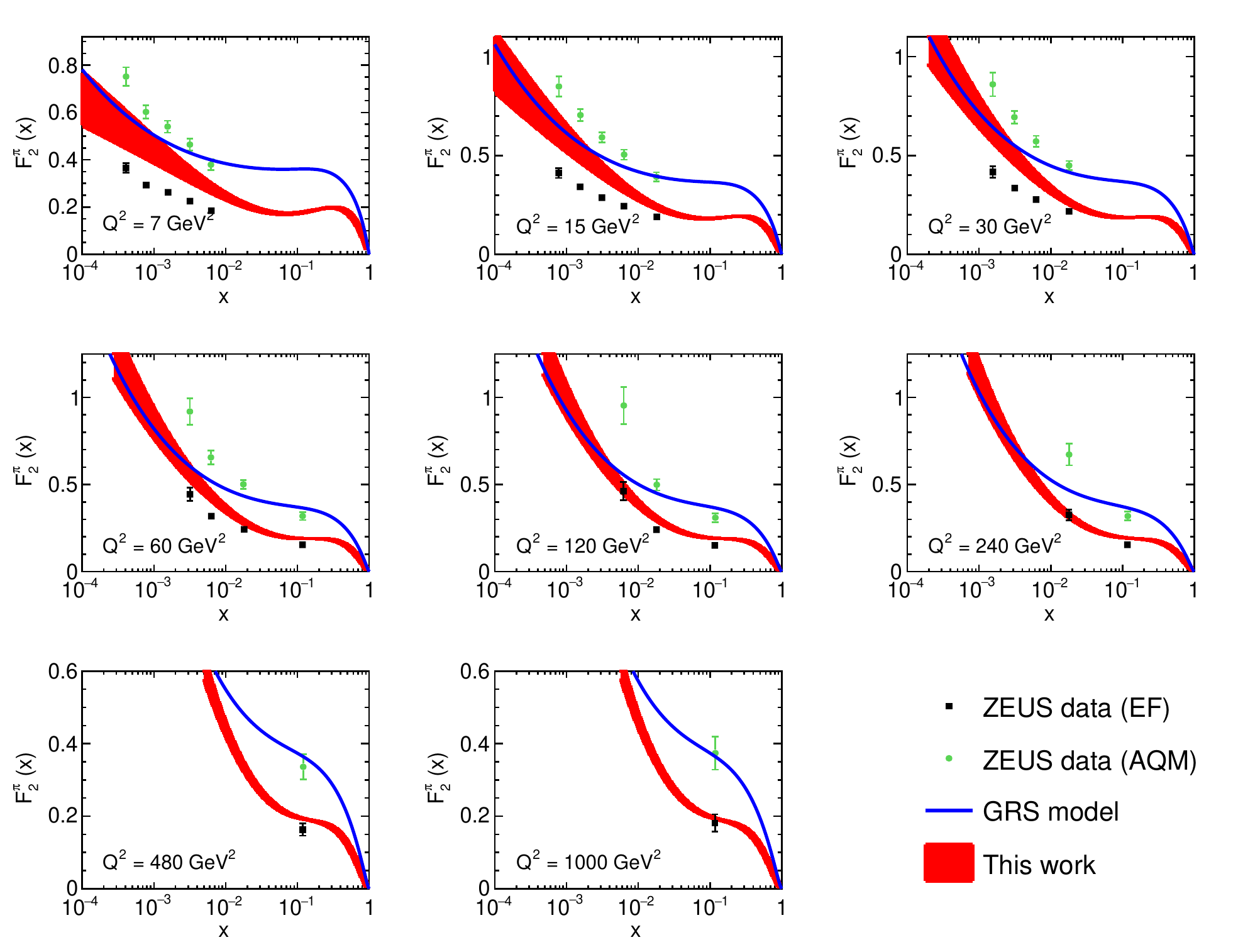}
\caption{Our predicted structure function $F_{2}^{\pi}(x, Q^{2})$ compared with the experimental data by ZEUS Collaboration \cite{Chekanov:2002pf},
and the GRS model predictions \cite{Gluck:1999xe}.
The pion structure function data are extracted using two different formulae (the EF model and the AQM model) \cite{Chekanov:2002pf}.
The error bands of this work show the 3$\sigma$ uncertainties of the structure function. }
\label{fig:F2pi-ZEUS}
\end{figure*}

By measuring leading neutron tagged deep inelastic scattering of $e$-$p$
and using one-pion exchange model, ZEUS and H1 Collaborations
at HERA extracted the pionic structure function at small $x$ \cite{Chekanov:2002pf,Aaron:2010ab}.
Fig. \ref{fig:F2pi-H1} shows the comparisons among the obtained pionic structure function
at small $x$, the H1 experimental data \cite{Aaron:2010ab}, and the GRS model predictions \cite{Gluck:1999xe}.
In the sea quark region, our global analysis with the MEM nonperturbative input agrees well with the experimental observations.
The previous GRS analysis is also close to the H1 data,
which is a pioneering work using also the dynamical parton model \cite{Gluck:1999xe}.
Fig. \ref{fig:F2pi-ZEUS} shows the comparisons among the obtained pionic structure function
at small $x$, the ZEUS experimental data \cite{Chekanov:2002pf}, and the GRS model predictions \cite{Gluck:1999xe}.
In the analysis by ZEUS Collaboration, the pion flux in LN-DIS
is evaluated by the effective field (EF) formula or the addictive quark model (AQM) formula \cite{Chekanov:2002pf}.
Both our result and GRS result are between the two experimental data sets from two different analysis methods.
The pionic structure function in this work is closer to the ZEUS data applying the EF method.
Note that the ZEUS data are not included in our global fit.
It is very interesting to find that our prediction is similar to the GRS prediction
in the region of $x<10^{-3}$, while there is big discrepancy between our prediction and the GRS prediction
around $x=0.1$. The future experiments on electron-ion colliders will
help clarify the differences among different models.

\begin{figure}[htbp]
\centering
\includegraphics[scale=0.41]{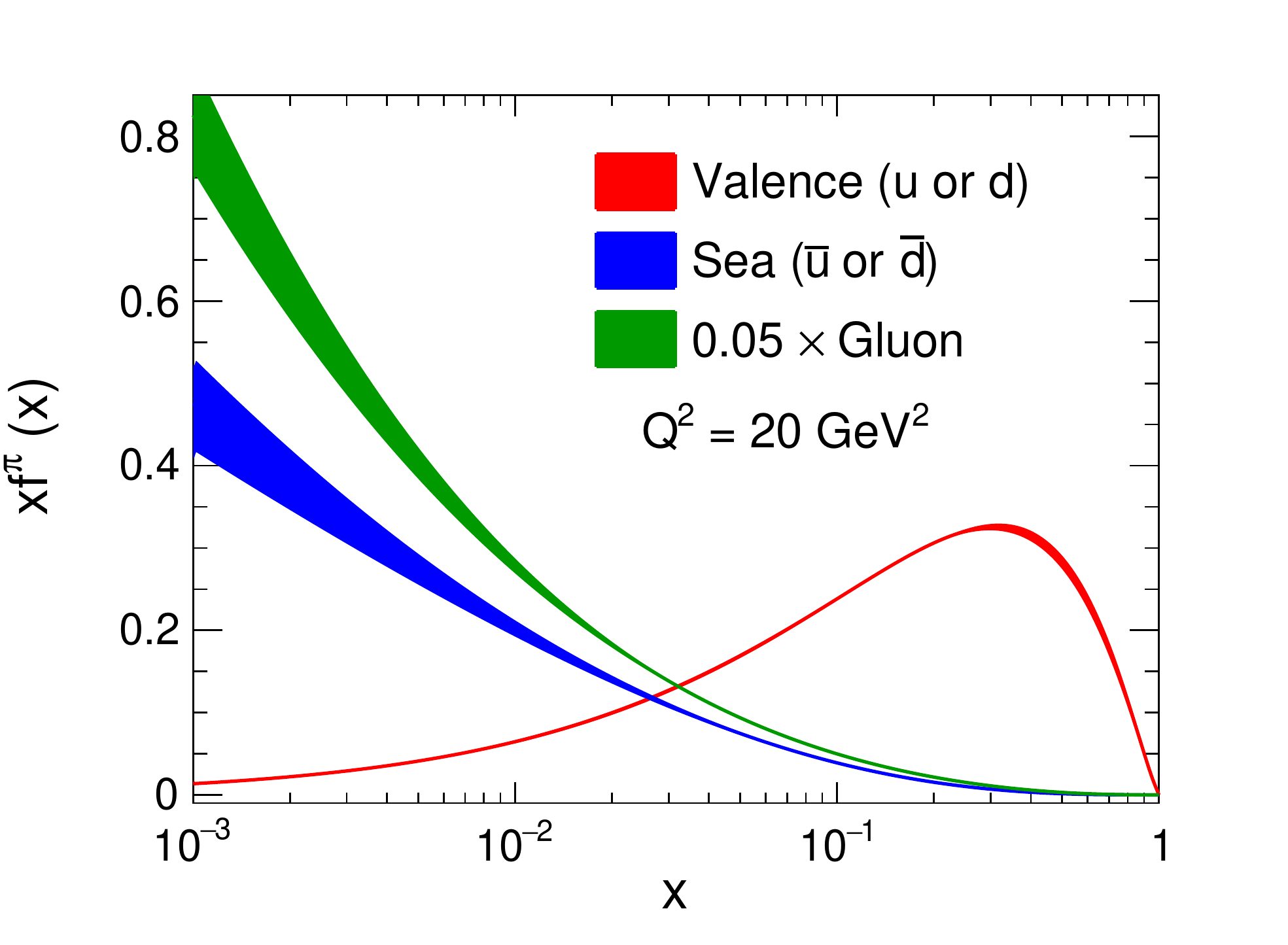}
\caption{The valence quark, sea quark and gluon distributions of the pion from this work.
The error bands show the 3$\sigma$ uncertainties of the quantities.  }
\label{fig:pionPDFs}
\end{figure}

\begin{figure}[htbp]
\centering
\includegraphics[scale=0.41]{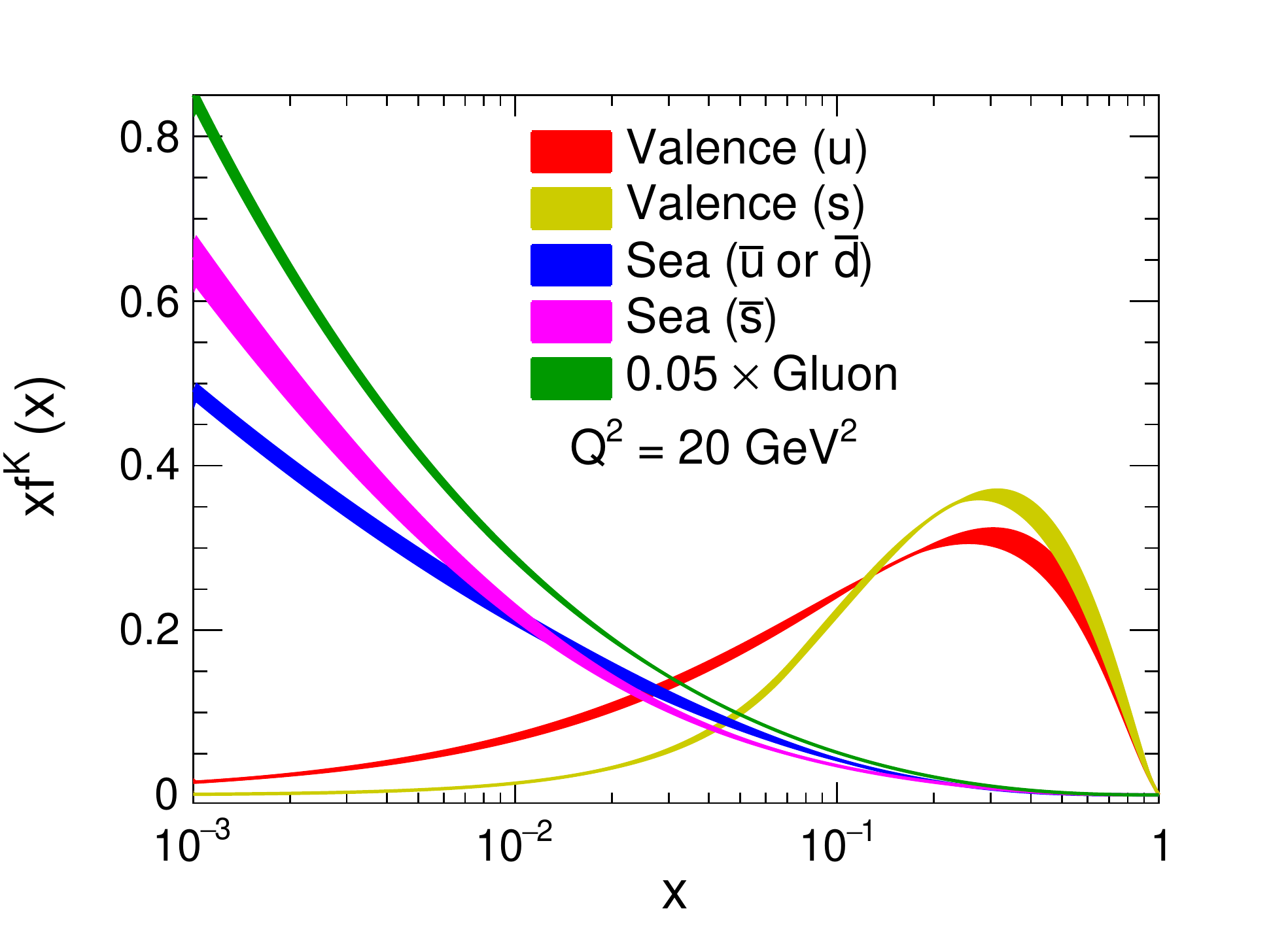}
\caption{The valence quark, sea quark and gluon distributions of the kaon from this work.
The error bands show the 3$\sigma$ uncertainties of the quantities. }
\label{fig:kaonPDFs}
\end{figure}

Fig. \ref{fig:pionPDFs} shows the obtained valence quark, sea quark and gluon distributions
of the pion at high $Q^2$ based on this global analysis. Fig. \ref{fig:kaonPDFs} shows the obtained valence quark,
sea quark and gluon distributions of the kaon at high $Q^2$ based on this global analysis.
In this analysis, the nonperturbative input and the parton-parton correlation length $R$
are the same for pion and kaon. The PDFs of the pion and the kaon exhibit very small differences
since there is only a little bit difference between the input hadronic scales for the pion and the kaon.

\begin{figure}[htbp]
\centering
\includegraphics[scale=0.41]{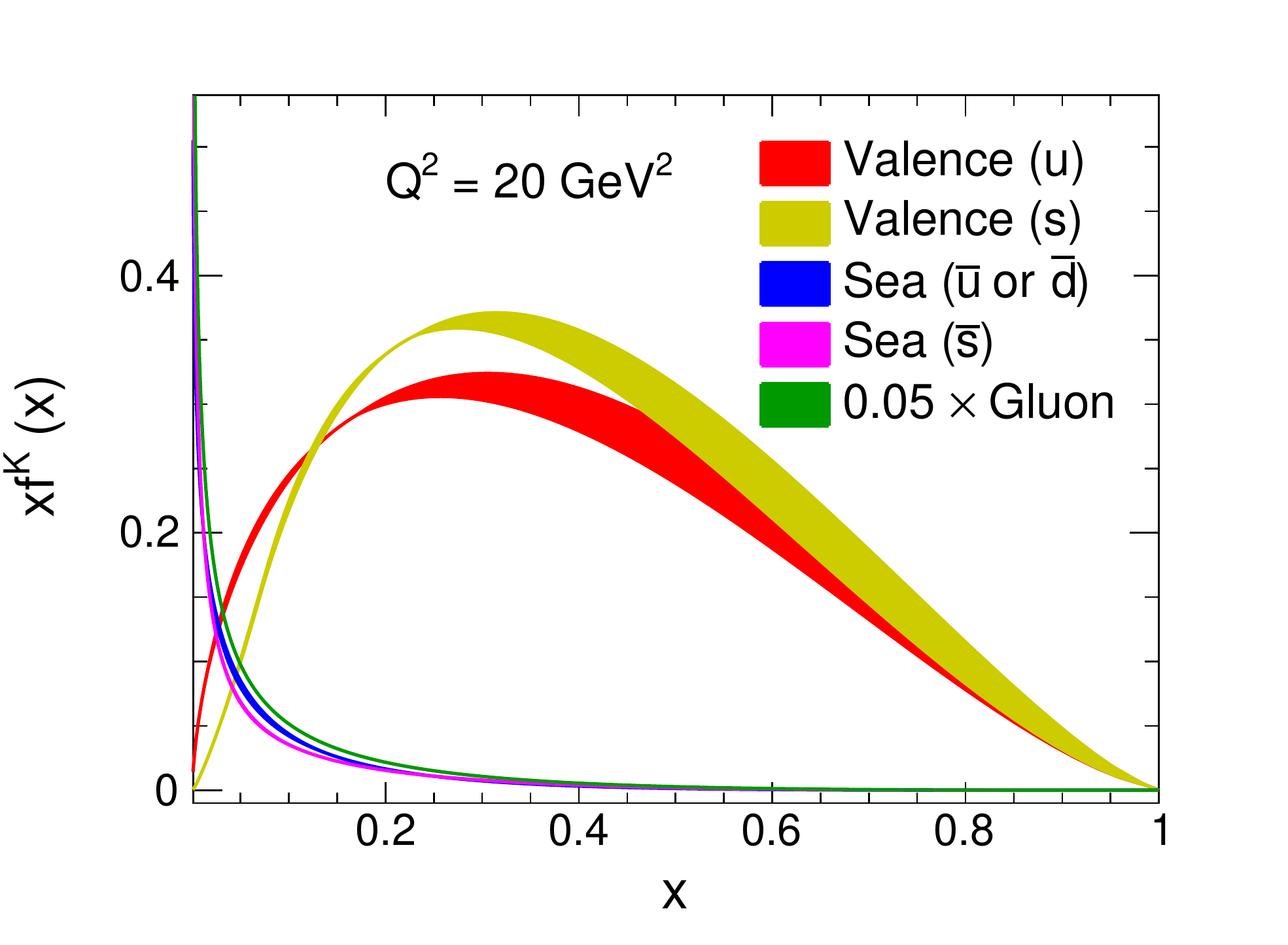}
\caption{The valence quark, sea quark and gluon distributions of the kaon from this work,
in the range of large $x$. The error bands show the 3$\sigma$ uncertainties of the quantities.   }
\label{fig:kaonPDFs-largeX}
\end{figure}

\begin{figure}[htbp]
\centering
\includegraphics[width=0.41\textwidth]{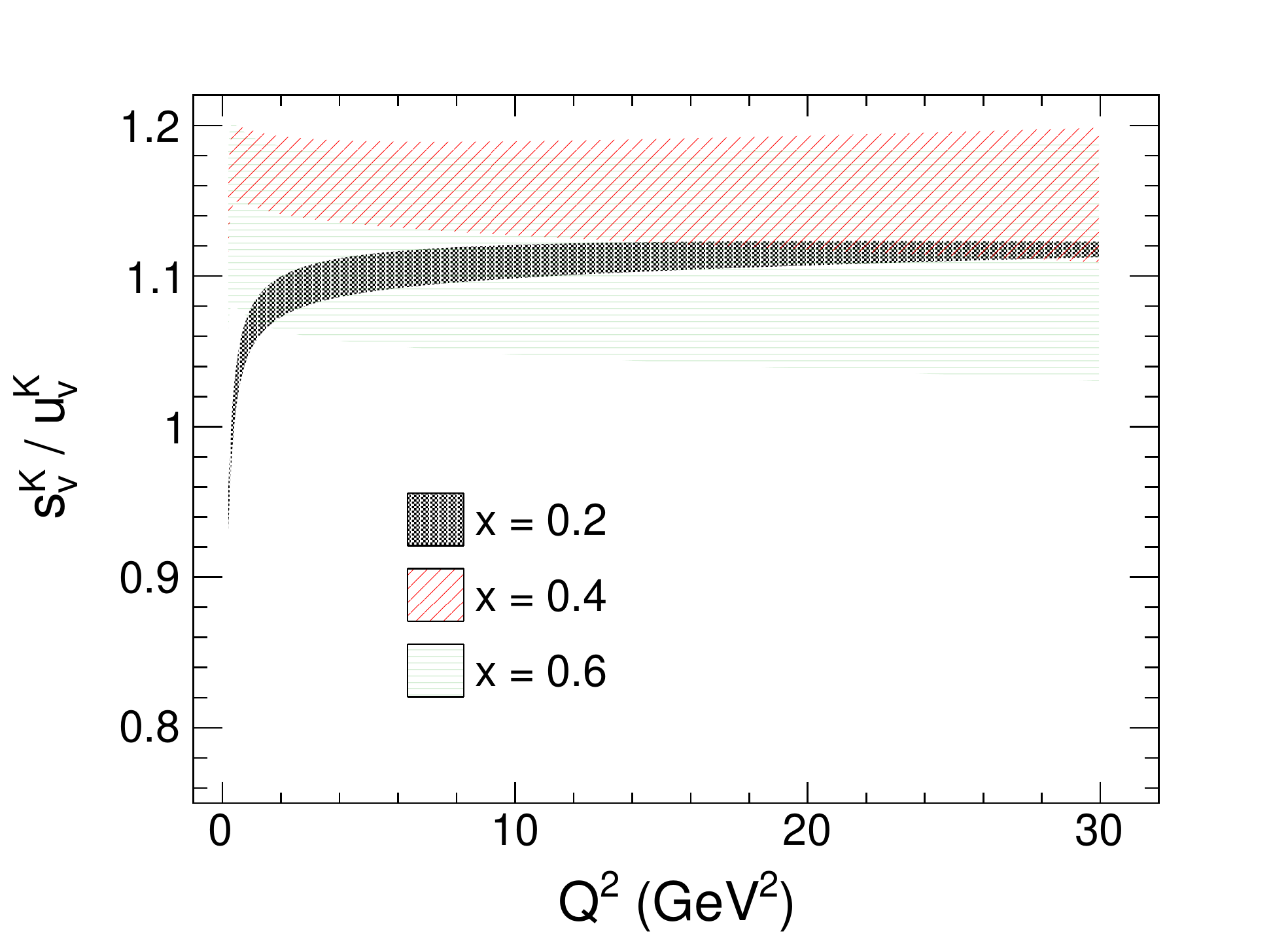}
\caption{The obtained ratio $s^{\rm K}_{\rm v} / u^{\rm K}_{\rm v}$ as a function of $Q^{2}$
with the strange quark mass effect is considered in the parton splitting processes of the DGLAP evolution.
The error bands show the 3$\sigma$ uncertainties of the ratio.  }
\label{fig:sv-over-uv-kaon}
\end{figure}

Fig. \ref{fig:kaonPDFs-largeX} demonstrates the obtained kaon PDFs in the large-$x$ region.
We find that the valence up quark distribution and the valence anti-strange quark distribution
are obviously different in the kaon. This is due to the mass-dependent splitting kernels
for our QCD evolution. Similar to a recent work \cite{Cui:2020dlm}, we apply a modified QCD splitting kernel function
for the strange quark distribution evolution starting from a very low $Q_0^2$,
since strange quark is significantly heavier than up or down quark.
The strange quark has a smaller probability of radiating the gluons,
hence the valence strange quark distribution is higher than the valence up quark distribution.
Fig. \ref{fig:sv-over-uv-kaon} shows the ratio of strange valence distribution
to up valence distribution as a function of $Q^2$.
The pure mass effect of the strange quark splitting is obvious but not huge.
In the large-$x$ region, the strange quark distribution is just around 15\% higher
than the up quark distribution.

\begin{figure*}[htbp]
\centering
\includegraphics[scale=0.9]{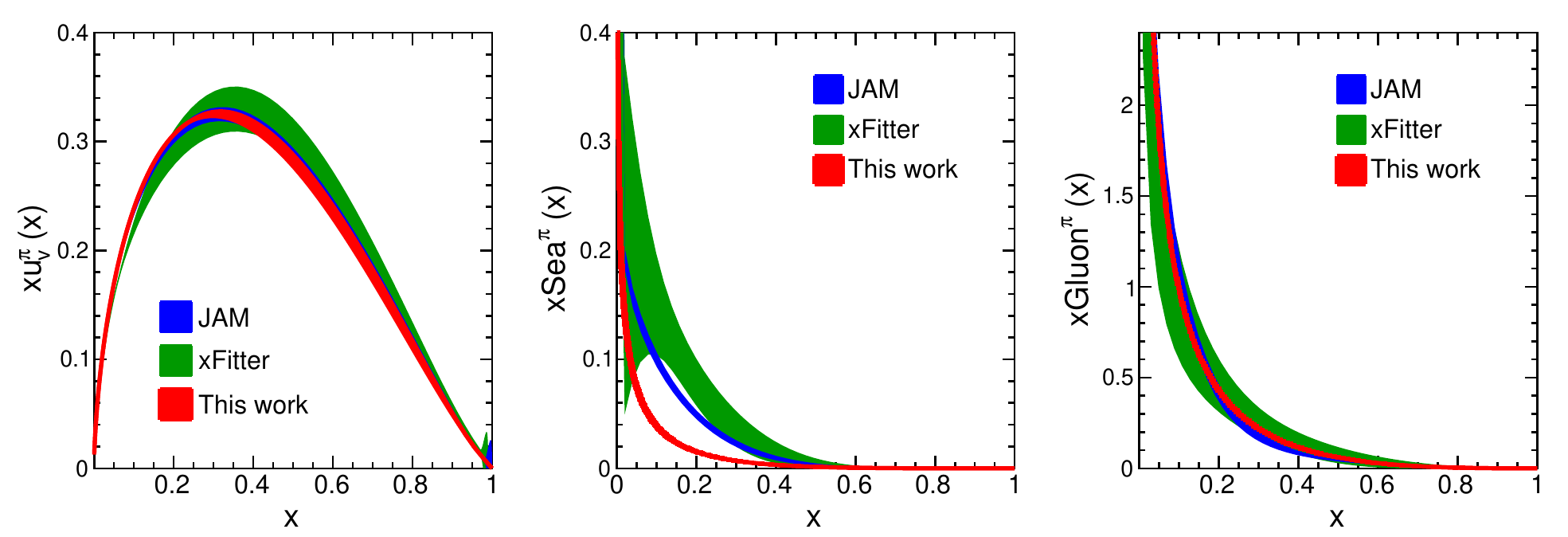}
\caption{Our obtained valence quark, sea quark, and gluon distributions of the pion
compared with the recent analyses from JAM18 \cite{Barry:2018ort} and xFitter \cite{Novikov:2020snp}, at $Q^2=20$ GeV$^2$.
The error bands of this work show the 3$\sigma$ uncertainties of the quantities.
}
\label{fig:pionPDF-sets}
\end{figure*}

Fig. \ref{fig:pionPDF-sets} shows the comparisons among our determined pion
PDFs with the popular pion PDFs from JAM analysis \cite{Barry:2018ort} and xFitter analysis \cite{Novikov:2020snp}.
Amazingly, we see the good agreements among our analysis and others' analyses
for valence quark and gluon distributions.
However our obtained dynamical sea quark distribution of the pion is lower than that from others
in the region of $x\gtrsim 0.01$.
We guess there are probably some intrinsic light sea quarks inside the pion
in addition to the pure dynamical sea quarks from the QCD evolution.
At small $x$, the dynamical sea quark distribution is consistent with JAM's
and xFitter's results.

The large-$x$ behavior of valence quark distribution is an interesting topic in theory,
which can be calculated with LQCD \cite{Sufian:2020vzb,Joo:2019bzr,Gao:2020ito},
quark number counting rule \cite{Brodsky:1973kr,Ball:2016spl},
and perturbative QCD predictions \cite{Ezawa:1974wm,Farrar:1975yb,Berger:1979du}.
The large-$x$ behavior is usually described by,
\begin{equation}
\begin{split}
x\rightarrow 1 : xf_{\rm i}(x,Q^2) \propto (1-x)^{\beta_{\rm i}},
\end{split}
\label{eq:largeX-behavior}
\end{equation}
with the parton distribution going down as the power $\beta$ of $(1-x)$.
With the obtained PDFs of pion and kaon at high $Q^2$, we performed the fits to them
and extracted the corresponding powers $\beta$, which are listed in table \ref{tab:beta-largeX}.
In the fits, we use the function form $C(1-x)^{\beta}$, and the fit range is $0.6<x<1$.
The errors in the table are propagated from the uncertainties of the parameters $Q_0^2$ and $R$,
rather than from the uncertainty of the model fit.
The coefficients $\beta$ are around 1 for the valence quark distributions;
The coefficients $\beta$ are around 4 for the sea quark distributions;
And the coefficients $\beta$ are around 3 for the gluon distributions.
At large $x$, the Brodsky-Farrar quark counting rule predicts
that $xf_{\rm i} \sim (1-x)^{2n_{\rm s} - 1}$, where $n_{\rm s}$ is the minimum number
of spectator partons \cite{Brodsky:1973kr,Ball:2016spl}.
Our results are consistent with the counting rule for the valence quark distribution and the gluon distribution.
The recent LQCD calculation based on LaMET (Large Momentum Effective Theory)
framework predicts $\beta_{\rm v}=1.53_{-(21)(25)}^{+(21)(25)}$ at $Q^2=10$ GeV$^2$ \cite{Gao:2020ito}.
Another recent LQCD calculation using the reduced Ioffe-time pseudodistributions
predicts $\beta_{\rm v}=1.08(41)(11)$ at $Q^2=4$ GeV$^2$ \cite{Joo:2019bzr}.
However $\beta_{\rm v}=2.12(56)(14)$ at $Q^2=4$ GeV$^2$ is predicted
from the current-current correlation in LQCD \cite{Sufian:2020vzb}.
From perturbative QCD, $\beta$ for the valence quark distribution
is 2 in the asymptotic region \cite{Ezawa:1974wm,Farrar:1975yb,Berger:1979du}.
Our extracted value in LO analysis is smaller than 2.
The discrepancy should be understood with the NLO corrections or the resummation of soft gluons near threshold \cite{Aicher:2010cb}.
L. Chang et al. \cite{Chang:2020kjj} point out that the naive quark number counting rule \cite{Brodsky:1973kr,Ball:2016spl}
should be modified for the hadron of zero spin.
In short, the large-$x$ behavior of the obtained valence quark distribution is between the quark number counting rule
and the perturbation QCD predictions.

\begin{table}
  \caption{
    Assuming PDFs obey $\propto (1-x)^{\beta}$ in the large-$x$ region approaching 1,
    we list the obtained values of the power $\beta$ for valence quark, sea quark and gluon distributions of
    the pion and the kaon at $Q^2=20$ GeV$^2$.
    The $\beta$ values are extracted from the fits in the range of $0.6<x<1$.
    The errors in the table indicate the 1$\sigma$ uncertainties.
  }
  \label{tab:beta-largeX}
  \begin{tabular}{cccccc}
  \hline\hline
            & $\beta_{\rm u_{v}}$ & $\beta_{\rm \bar{s}_{v}}$ & $\beta_{\rm \bar{u}_{sea}}$ & $\beta_{\rm \bar{s}_{sea}}$ & $\beta_{\rm g}$  \\
  \hline
    $\pi$   &   1.09(5)            &      -                    &   4.43(2)                     &               4.43(2)        & 3.03(47)      \\
    K       &   1.19(5)            &      1.25(5)              &   4.19(14)                    &               3.94(4)        & 3.14(50)      \\
  \hline\hline
  \end{tabular}
\end{table}

\begin{table}
  \caption{
    Assuming PDFs obey $\propto x^{\alpha}$ in the small-$x$ region approaching 0,
    we list the obtained values of the power $\alpha$ for valence quark, sea quark and gluon distributions of
    the pion and the kaon at $Q^2=20$ GeV$^2$.
    The $\alpha$ values are extracted from the fits in the range of $10^{-6}<x<10^{-1}$.
    The errors in the table indicate the 1$\sigma$ uncertainties.
  }
  \label{tab:alpha-smallX}
  \begin{tabular}{cccccc}
  \hline\hline
            & $\alpha_{\rm u_{v}}$ & $\alpha_{\rm \bar{s}_{v}}$ & $\alpha_{\rm \bar{u}_{sea}}$ & $\alpha_{\rm \bar{s}_{sea}}$ & $\alpha_{\rm g}$  \\
  \hline
    $\pi$   &   0.59(1)               &      -                     &        -0.19(1)           &            -0.19(1)          & -0.25(1)     \\
    K       &   0.57(1)               &      1.19(3)               &        -0.20(1)           &            -0.21(1)          & -0.25(1)    \\
  \hline\hline
  \end{tabular}
\end{table}

In small-$x$ region, the $x$-dependencies of quark and gluon distributions are described
well by Regge theory \cite{Ball:2016spl,Regge:1959mz}. The small-$x$ behavior is usually described by,
\begin{equation}
\begin{split}
x\rightarrow 0 : xf_{\rm i}(x,Q^2) \propto x^{\alpha_{\rm i}}.
\end{split}
\label{eq:smallX-behavior}
\end{equation}
With the obtained PDFs of pion and kaon at high $Q^2$, we performed the fits to them
and extracted the corresponding powers $\alpha$, which are listed in table \ref{tab:alpha-smallX}.
The exponents $\alpha$ are around 0.5 for the valence quark distributions;
The exponent $\alpha$ is around 1 for the strange valence quark distribution;
The exponents $\alpha$ are around -0.2 for the sea quark distributions;
And the exponents $\alpha$ are around -0.25 for the gluon distributions.
In the fits, we use the function form $Cx^{\alpha}$, and the fit range is $10^{-6}<x<0.1$.
The Regge theory \cite{Ball:2016spl,Regge:1959mz} predicts $\alpha=0.5$ for valence quark distributions,
$\alpha=-0.08$ for sea quark and gluon distributions.
With the resummation of double logarithms, the Regge theory then gives
$\alpha=0.63$ for valence quark distributions, and $\alpha=-0.2$ for sea quark and gluon distributions.
The state-of-the-art LQCD calculation based on LaMET predicts $\alpha_{\rm v}=0.45_{-(08)(09)}^{+(11)(09)}$ at $Q^2=10$ GeV$^2$ \cite{Gao:2020ito}.
Another recent LQCD calculation based on pseudo-PDFs predicts $\alpha_{\rm v}=0.52(14)(4)$ at $Q^2=4$ GeV$^2$ \cite{Joo:2019bzr}.
And the analysis from current-current correlation in LQCD \cite{Sufian:2020vzb}
predicts $\alpha_{\rm v}=0.78(11)(3)$ at $Q^2=4$ GeV$^2$.
The PDFs of the pion and the kaon determined in this work are consistent with
the predictions of Regge theory \cite{Ball:2016spl,Regge:1959mz}
and LQCD approach \cite{Sufian:2020vzb,Joo:2019bzr,Gao:2020ito}.

The moments of quark distributions are usually calculated in theory.
To compare with the theoretical calculations, we also provide the moments of the obtained
parton distributions of the pion and the kaon.
The moment of the parton momentum fraction distribution used in this work is defined as,
\begin{equation}
\left<x^{n}\right>_{f} = \int_0^1 x^{n} f(x,Q^2) dx.
\label{MomentumSum}
\end{equation}
The first three moments of the quark distributions of the pion are listed in table \ref{tab:PionMoments}.
The first three moments of the quark distributions of the kaon are listed in table \ref{tab:KaonMoments}.
LQCD and DSE are two main nonperturbative QCD approaches in understanding the strong interaction.
From the tables, we can see that the PDF moments of the pion and the kaon determined in this analysis
are consistent with QCD theory.
The DSE result is taken from the recent work by Z. F. Cui et al. \cite{Chen:2016sno,Cui:2020dlm},
and the LQCD results are taken from Refs. \cite{Best:1997qp,Detmold:2003tm,Lin:2020ssv,Joo:2019bzr,Gao:2020ito}.

\begin{table}
  \caption{
    Comparisons of our predicted moments $\left<x\right>$, $\left<x^2\right>$, $\left<x^3\right>$ of
    up valence quark distribution of the pion
    with some other theoretical calculations, at $Q^2=4,~10,~27$ GeV$^2$.
    The errors in the table indicate the 1$\sigma$ uncertainties.
  }
  \label{tab:PionMoments}
  \begin{tabular}{ccccc}
  \hline\hline
         & $\left<x\right>^{\rm\pi}$  & $\left<x^2\right>^{\rm\pi}$   & $\left<x^3\right>^{\rm\pi}$  & $Q^2$/GeV$^2$  \\
  \hline
    DSE \cite{Chen:2016sno}    &           0.26(8)       &         0.11(4)        &      0.058(27)   &  4 \\
    LQCD \cite{Detmold:2003tm}   &         0.24(1)       &         0.09(3)        &      0.043(15)   &  4 \\
    LQCD \cite{Best:1997qp}   &            0.273(12)     &         0.107(35)      &      0.048(20)   &  4 \\
    LQCD \cite{Gao:2020ito}   &            0.213(19)     &         0.101(68)      &      0.061(40)   &  10 \\
    LQCD \cite{Joo:2019bzr}   &            0.165(9)      &         0.064(1)       &      0.033(2)    &  27 \\
    LFHQCD \cite{deTeramond:2018ecg}   &   0.233         &         0.102          &      0.056    & 4  \\
  NJL model \cite{Davidson:1994uv}     &   0.236         &         0.103          &      0.057    & 4 \\
  JAM                &            0.243(5)      &       0.109(3)       &      0.061(2)  &  4   \\
  xFitter                    &            0.24(2)      &         0.11(1)       &      0.063(4)  &  4   \\
  This work                    &            0.241(3)      &         0.107(2)       &      0.060(2)  &  4   \\
  This work                    &            0.221(3)      &         0.093(2)       &      0.050(1)  &  10   \\
  This work                    &            0.205(3)      &         0.083(2)       &      0.043(1)  &  27   \\
  \hline\hline
  \end{tabular}
\end{table}

\begin{table}
  \caption{
    Comparisons of our predicted moments $\left<x\right>$, $\left<x^2\right>$, $\left<x^3\right>$ of
    quark distributions of the kaon
    with some other theoretical calculations, at $Q^2=27$ GeV$^2$.
    The errors in the table indicate the 1$\sigma$ uncertainties.
  }
  \label{tab:KaonMoments}
  \begin{tabular}{cccc}
  \hline\hline
         & $\left<x\right>^{\rm K}$  & $\left<x^2\right>^{\rm K}$   &  $\left<x^3\right>^{\rm K}$    \\
  \hline
    DSE \cite{Cui:2020dlm} ($u_{\rm v}$)   &        0.19(2)            &           0.067(9)      &         0.030(5)    \\
    DSE \cite{Cui:2020dlm} ($\bar{s}_{\rm v}$)   &     0.23(2)         &           0.085(11)      &         0.040(7)    \\
    LQCD \cite{Lin:2020ssv} ($u_{\rm v}$)   &          0.192(8)        &           0.080(7)      &         0.041(6)    \\
    LQCD \cite{Lin:2020ssv} ($\bar{s}_{\rm v}$)   &    0.261(8)        &           0.120(7)      &         0.069(6)    \\
  This work ($u_{\rm v}$)         &           0.192(6)              &           0.0747(32)   &      0.0382(21)    \\
  This work ($\bar{s}_{\rm v}$)   &           0.207(5)              &           0.0828(31)   &      0.0422(21)    \\
  \hline\hline
  \end{tabular}
\end{table}

\begin{figure}[htbp]
\centering
\includegraphics[width=0.41\textwidth]{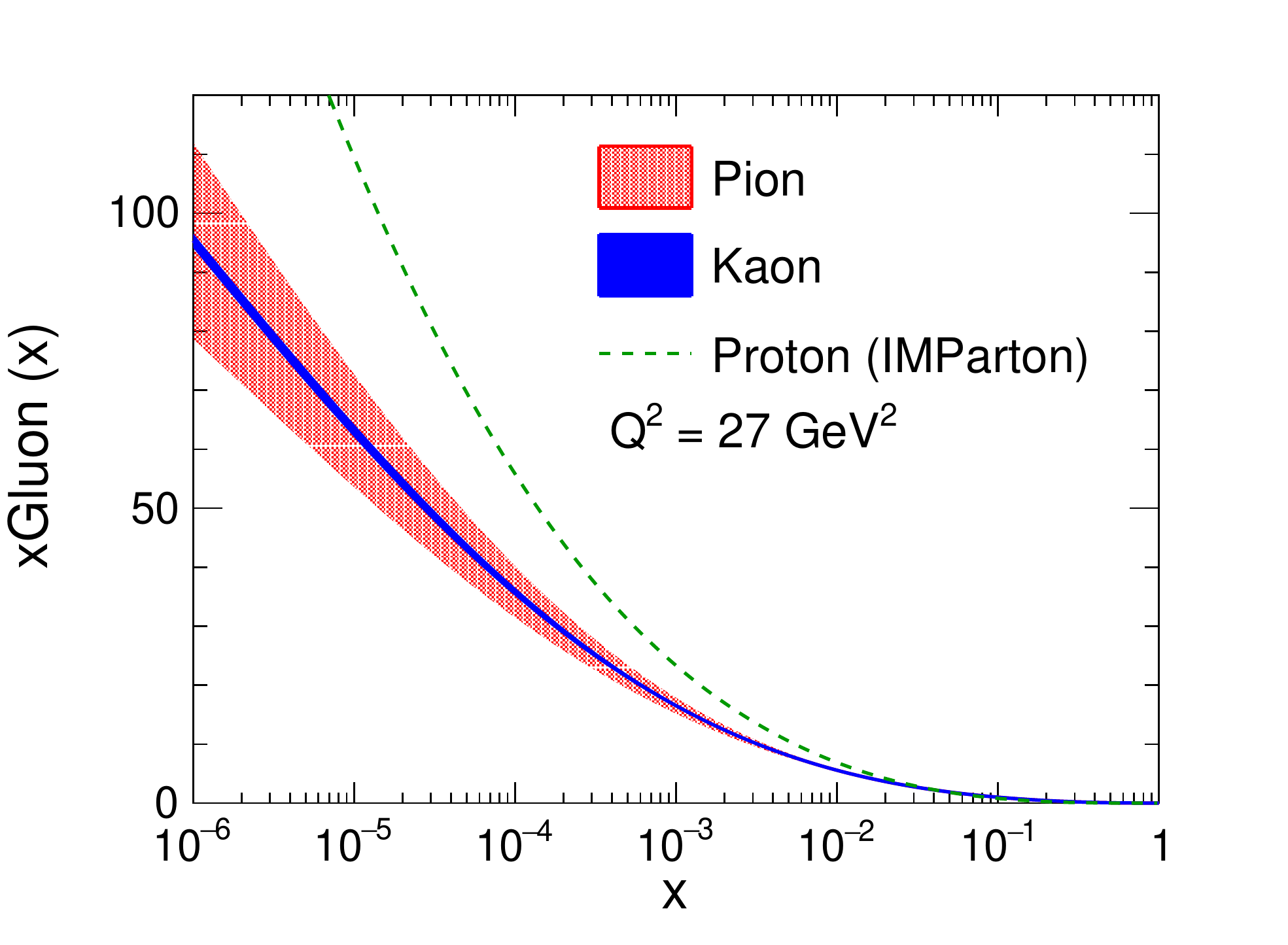}
\caption{
  The comparisons of the gluon distributions inside the pion, the kaon, and the proton at $Q^2=27$ GeV$^2$.
  The error bands of this work show the 3$\sigma$ uncertainties of the quantities.   }
\label{fig:gluon-comparisons}
\end{figure}

\begin{figure}[htbp]
\centering
\includegraphics[width=0.42\textwidth]{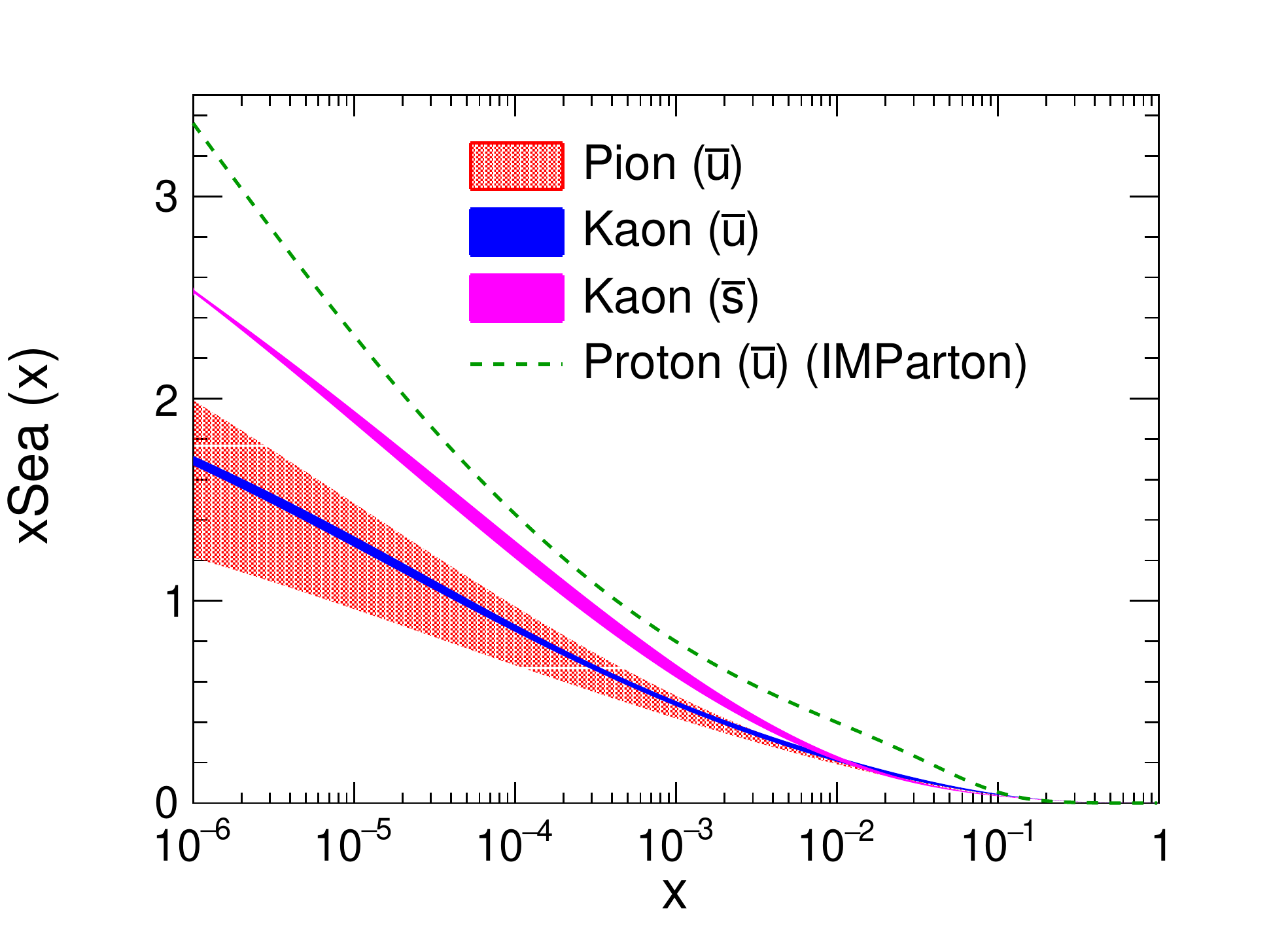}
\caption{
  The comparisons of the sea quark distributions inside the pion, the kaon, and the proton at $Q^2=27$ GeV$^2$.
  The error bands of this work show the 3$\sigma$ uncertainties of the quantities.   }
\label{fig:sea-comparisons}
\end{figure}

Fig. \ref{fig:gluon-comparisons} shows the gluon momentum distributions of the pion, the kaon and the proton (IMParton \cite{Wang:2016sfq})
at a high $Q^2$. We find that the gluon distributions of the pion and the kaon are quite similar,
which are lower than the gluon distribution of the proton at small $x$.
Fig. \ref{fig:sea-comparisons} shows the sea quark momentum distributions of the pion, the kaon and the proton (IMParton \cite{Wang:2016sfq})
at a high $Q^2$. We find that the sea quark distributions of the pion and the kaon are quite similar,
which are lower than the sea quark distribution of the proton at small $x$ as well.
Here we take the proton PDFs of IMParton as the reference, and this is because
the IMParton PDFs of the proton are also determined under the dynamical parton model,
where the gluon distribution is completely produced from QCD evolution equations.
The other finding is that the strange sea quark distribution is higher than the up sea quark distribution
in the kaon. This is due to the modified splitting kernels used for the kaon in the DGLAP evolution.

\section{Summary}
\label{sec:summary}

In this analysis, we have obtained the hadronic scales for the pion and the kaon,
where there are only the valence components of the hadrons at the scale.
The input hadronic scales for the pion and the kaon are around 0.1 GeV$^2$,
which are slightly higher than that for the proton \cite{Wang:2016sfq}.
In our analysis, the input parton distributions of merely valence quarks
are the suggested uniform distribution by a maximum entropy estimation.
The obtained parton distributions of the pion and the kaon are shown
to be consistent with the DY data from E615 and the LN-DIS
from H1 and ZEUS at HERA.
Via this analysis, we can see that the nonperturbative input from MEM is reasonable.
For the practical applications based on this analysis,
we provide a software package for the users to access the obtained PDFs of the pion and the kaon,
which can be found on the web \cite{piIMParton-github}.

In this work, we use the DGLAP equations with parton-parton recombination corrections.
By the global fit to the current experimental data, we get the parton-parton correlation length $R$
for the pion and the kaon, which is smaller than that for the proton.
This is within the naive expectation that the parton-parton correlation length
is proportional to the size of the hadron.
Since the radii of the pion and the kaon are smaller than that of the proton,
it is natural to have the two-parton fusion probability higher inside the pion and the kaon.
The QCD evolution with the nonlinear parton-parton recombination corrections is an approximate but
simple bridge to connect the pure valence nonperturbative input with the experimental measurements at the hard scales.

The purely dynamical parton model is used in this work.
In this dynamical parton model, there are no sea quarks and gluons
at the input hadronic scale. Therefore, the sea quark and gluon distributions
at high scales are totally generated in the standard QCD evolution.
Although there is the model-dependence,
the sea quark and gluon distributions are strictly constrained.

The large-$x$ and small-$x$ behaviors of the obtained PDFs
of the pion and the kaon are studied. We find that the power of $(1-x)$ at large $x$ is between
the quark number counting rule and the perturbative QCD predictions.
And the power law behavior at small $x$ is in well agreement with Regge theory
calculations with the double logarithmic resummation considered.
We also calculate the first three leading moments $\left<x\right>$, $\left<x^2\right>$, $\left<x^3\right>$,
and compare them with LQCD calculations and other theoretical calculations.
The consistencies among them are found.

For this analysis, we also see that
there are some rooms left to be improved in the future, such as using the DGLAP equations at
next-to-leading order with the parton-parton recombination corrections.
Most importantly, much more experimental data of the pion and the kaon are expected
to constrain the sea quark and gluon distributions of the mesons.
Especially, the kaon structure experiment is highly needed,
as there are few data on the kaon structure for now.
In the future, the updating and the new facilities will provide much more data
in settling down the questions, such as the COMPASS++/AMBER with the pion and kaon beams \cite{Denisov:2018unj},
JLab \cite{JLabPionProposal}, EicC \cite{Chen:2020ijn,Chen:2018wyz} and EIC \cite{Accardi:2012qut}
with the tagged DIS techniques.

\section*{Acknowledgments}
This work is supported by the Strategic Priority Research Program of Chinese Academy of Sciences
under the Grant NO. XDB34030301.

\bibliographystyle{unsrt}
\bibliography{refs}{}

\begin{thebibliography}{10}

\bibitem{Nambu:1960tm}
Yoichiro Nambu.
\newblock {Quasiparticles and Gauge Invariance in the Theory of
  Superconductivity}.
\newblock {\em Phys. Rev.}, 117:648--663, 1960.

\bibitem{Goldstone:1962es}
Jeffrey Goldstone, Abdus Salam, and Steven Weinberg.
\newblock {Broken Symmetries}.
\newblock {\em Phys. Rev.}, 127:965--970, 1962.

\bibitem{Maris:1997hd}
Pieter Maris, Craig~D. Roberts, and Peter~C. Tandy.
\newblock {Pion mass and decay constant}.
\newblock {\em Phys. Lett. B}, 420:267--273, 1998.

\bibitem{Roberts:2019ngp}
Craig~D Roberts.
\newblock {Insights into the Origin of Mass}.
\newblock In {\em {27th International Nuclear Physics Conference}}, 9 2019.

\bibitem{Roberts:2020udq}
Craig~D. Roberts and Sebastian~M. Schmidt.
\newblock {Reflections upon the emergence of hadronic mass}.
\newblock {\em Eur. Phys. J. ST}, 229(22-23):3319--3340, 2020.

\bibitem{Chen:2020ijn}
Xurong Chen, Feng-Kun Guo, Craig~D. Roberts, and Rong Wang.
\newblock {Selected Science Opportunities for the EicC}.
\newblock {\em Few Body Syst.}, 61(4):43, 2020.

\bibitem{Ji:1994av}
Xiang-Dong Ji.
\newblock {A QCD analysis of the mass structure of the nucleon}.
\newblock {\em Phys. Rev. Lett.}, 74:1071--1074, 1995.

\bibitem{Yang:2018nqn}
Yi-Bo Yang, Jian Liang, Yu-Jiang Bi, Ying Chen, Terrence Draper, Keh-Fei Liu,
  and Zhaofeng Liu.
\newblock {Proton Mass Decomposition from the QCD Energy Momentum Tensor}.
\newblock {\em Phys. Rev. Lett.}, 121(21):212001, 2018.

\bibitem{Lorce:2017xzd}
C\'edric Lorc\'e.
\newblock {On the hadron mass decomposition}.
\newblock {\em Eur. Phys. J. C}, 78(2):120, 2018.

\bibitem{Chen:2020gml}
Ying Chen.
\newblock {The Origin of Hadron Masses}.
\newblock 10 2020.

\bibitem{Cui:2020dlm}
Zhu-Fang Cui, Minghui Ding, Fei Gao, Kh\'epani Raya, Daniele Binosi, Lei Chang,
  Craig~D Roberts, Jose Rodr\'\i{}guez-Quintero, and Sebastian~M Schmidt.
\newblock {Higgs modulation of emergent mass as revealed in kaon and pion
  parton distributions}.
\newblock {\em Eur. Phys. J. A}, 57(1):5, 2021.

\bibitem{Detmold:2003tm}
William Detmold, W.~Melnitchouk, and Anthony~William Thomas.
\newblock {Parton distribution functions in the pion from lattice QCD}.
\newblock {\em Phys. Rev. D}, 68:034025, 2003.

\bibitem{Shugert:2020tgq}
Charles Shugert, Xiang Gao, Taku Izubichi, Luchang Jin, Christos Kallidonis,
  Nikhil Karthik, Swagato Mukherjee, Peter Petreczky, Sergey Syritsyn, and Yong
  Zhao.
\newblock {Pion valence quark PDF from lattice QCD}.
\newblock In {\em {37th International Symposium on Lattice Field Theory}}, 1
  2020.

\bibitem{Sufian:2020vzb}
Raza~Sabbir Sufian, Colin Egerer, Joseph Karpie, Robert~G. Edwards, B\'alint
  Jo\'o, Yan-Qing Ma, Kostas Orginos, Jian-Wei Qiu, and David~G. Richards.
\newblock {Pion Valence Quark Distribution from Current-Current Correlation in
  Lattice QCD}.
\newblock {\em Phys. Rev. D}, 102(5):054508, 2020.

\bibitem{Joo:2019bzr}
B\'alint Jo\'o, Joseph Karpie, Kostas Orginos, Anatoly~V. Radyushkin, David~G.
  Richards, Raza~Sabbir Sufian, and Savvas Zafeiropoulos.
\newblock {Pion valence structure from Ioffe-time parton pseudodistribution
  functions}.
\newblock {\em Phys. Rev. D}, 100(11):114512, 2019.

\bibitem{Gao:2020ito}
Xiang Gao, Luchang Jin, Christos Kallidonis, Nikhil Karthik, Swagato Mukherjee,
  Peter Petreczky, Charles Shugert, Sergey Syritsyn, and Yong Zhao.
\newblock {Valence parton distribution of the pion from lattice QCD:
  Approaching the continuum limit}.
\newblock {\em Phys. Rev. D}, 102(9):094513, 2020.

\bibitem{Nguyen:2011jy}
Trang Nguyen, Adnan Bashir, Craig~D. Roberts, and Peter~C. Tandy.
\newblock {Pion and kaon valence-quark parton distribution functions}.
\newblock {\em Phys. Rev. C}, 83:062201, 2011.

\bibitem{Chen:2016sno}
Chen Chen, Lei Chang, Craig~D. Roberts, Shaolong Wan, and Hong-Shi Zong.
\newblock {Valence-quark distribution functions in the kaon and pion}.
\newblock {\em Phys. Rev. D}, 93(7):074021, 2016.

\bibitem{Ding:2019lwe}
Minghui Ding, Kh\'epani Raya, Daniele Binosi, Lei Chang, Craig~D Roberts, and
  Sebastian~M. Schmidt.
\newblock {Symmetry, symmetry breaking, and pion parton distributions}.
\newblock {\em Phys. Rev. D}, 101(5):054014, 2020.

\bibitem{deTeramond:2018ecg}
Guy~F. de~Teramond, Tianbo Liu, Raza~Sabbir Sufian, Hans~G\"unter Dosch,
  Stanley~J. Brodsky, and Alexandre Deur.
\newblock {Universality of Generalized Parton Distributions in Light-Front
  Holographic QCD}.
\newblock {\em Phys. Rev. Lett.}, 120(18):182001, 2018.

\bibitem{Lan:2019vui}
Jiangshan Lan, Chandan Mondal, Shaoyang Jia, Xingbo Zhao, and James~P. Vary.
\newblock {Parton Distribution Functions from a Light Front Hamiltonian and QCD
  Evolution for Light Mesons}.
\newblock {\em Phys. Rev. Lett.}, 122(17):172001, 2019.

\bibitem{Szczepaniak:1993uq}
A.~Szczepaniak, Chueng-Ryong Ji, and Stephen~R. Cotanch.
\newblock {Generalized relativistic meson wave function}.
\newblock {\em Phys. Rev. D}, 49:3466--3473, 1994.

\bibitem{Frederico:1994dx}
T.~Frederico and G.A. Miller.
\newblock {Deep inelastic structure function of the pion in the null plane
  phenomenology}.
\newblock {\em Phys. Rev. D}, 50:210--216, 1994.

\bibitem{Watanabe:2016lto}
Akira Watanabe, Chung~Wen Kao, and Katsuhiko Suzuki.
\newblock {Meson cloud effects on the pion quark distribution function in the
  chiral constituent quark model}.
\newblock {\em Phys. Rev. D}, 94(11):114008, 2016.

\bibitem{Badier:1980jq}
J.~Badier et~al.
\newblock {Measurement of the $K^- / \pi^-$ Structure Function Ratio Using the
  \{Drell-Yan\} Process}.
\newblock {\em Phys. Lett. B}, 93:354--356, 1980.

\bibitem{Badier:1983mj}
J.~Badier et~al.
\newblock {Experimental Determination of the pi Meson Structure Functions by
  the Drell-Yan Mechanism}.
\newblock {\em Z. Phys. C}, 18:281, 1983.

\bibitem{Betev:1985pf}
B.~Betev et~al.
\newblock {Differential Cross-section of High Mass Muon Pairs Produced by a
  194-\{GeV\}/$c \pi^-$ Beam on a Tungsten Target}.
\newblock {\em Z. Phys. C}, 28:9, 1985.

\bibitem{Conway:1989fs}
J.S. Conway et~al.
\newblock {Experimental Study of Muon Pairs Produced by 252-GeV Pions on
  Tungsten}.
\newblock {\em Phys. Rev. D}, 39:92--122, 1989.

\bibitem{Chekanov:2002pf}
S.~Chekanov et~al.
\newblock {Leading neutron production in e+ p collisions at HERA}.
\newblock {\em Nucl. Phys. B}, 637:3--56, 2002.

\bibitem{Aaron:2010ab}
F.D. Aaron et~al.
\newblock {Measurement of Leading Neutron Production in Deep-Inelastic
  Scattering at HERA}.
\newblock {\em Eur. Phys. J. C}, 68:381--399, 2010.

\bibitem{Gluck:1999xe}
M.~Gluck, E.~Reya, and I.~Schienbein.
\newblock {Pionic parton distributions revisited}.
\newblock {\em Eur. Phys. J. C}, 10:313--317, 1999.

\bibitem{Sutton:1991ay}
P.J. Sutton, Alan~D. Martin, R.G. Roberts, and W.James Stirling.
\newblock {Parton distributions for the pion extracted from Drell-Yan and
  prompt photon experiments}.
\newblock {\em Phys. Rev. D}, 45:2349--2359, 1992.

\bibitem{Barry:2018ort}
P.C. Barry, N.~Sato, W.~Melnitchouk, and Chueng-Ryong Ji.
\newblock {First Monte Carlo Global QCD Analysis of Pion Parton Distributions}.
\newblock {\em Phys. Rev. Lett.}, 121(15):152001, 2018.

\bibitem{Towell:2001nh}
R.S. Towell et~al.
\newblock {Improved measurement of the anti-d / anti-u asymmetry in the nucleon
  sea}.
\newblock {\em Phys. Rev. D}, 64:052002, 2001.

\bibitem{Novikov:2020snp}
Ivan Novikov et~al.
\newblock {Parton Distribution Functions of the Charged Pion Within The xFitter
  Framework}.
\newblock {\em Phys. Rev. D}, 102(1):014040, 2020.

\bibitem{Ezawa:1974wm}
Z.F. Ezawa.
\newblock {Wide-Angle Scattering in Softened Field Theory}.
\newblock {\em Nuovo Cim. A}, 23:271--290, 1974.

\bibitem{Farrar:1975yb}
Glennys~R. Farrar and Darrell~R. Jackson.
\newblock {Pion and Nucleon Structure Functions Near x=1}.
\newblock {\em Phys. Rev. Lett.}, 35:1416, 1975.

\bibitem{Berger:1979du}
Edmond~L. Berger and Stanley~J. Brodsky.
\newblock {Quark Structure Functions of Mesons and the Drell-Yan Process}.
\newblock {\em Phys. Rev. Lett.}, 42:940--944, 1979.

\bibitem{Chang:2020kjj}
Lei Chang, Kh\'epani Raya, and Xiaobin Wang.
\newblock {Pion Parton Distribution Function in Light-Front Holographic QCD}.
\newblock {\em Chin. Phys. C}, 44(11):114105, 2020.

\bibitem{Aicher:2010cb}
Matthias Aicher, Andreas Schafer, and Werner Vogelsang.
\newblock {Soft-gluon resummation and the valence parton distribution function
  of the pion}.
\newblock {\em Phys. Rev. Lett.}, 105:252003, 2010.

\bibitem{Gluck:1977ah}
M.~Gluck and E.~Reya.
\newblock {Dynamical Determination of Parton and Gluon Distributions in Quantum
  Chromodynamics}.
\newblock {\em Nucl. Phys. B}, 130:76--92, 1977.

\bibitem{Wang:2016sfq}
Rong Wang and Xurong Chen.
\newblock {Dynamical parton distributions from DGLAP equations with nonlinear
  corrections}.
\newblock {\em Chin. Phys. C}, 41(5):053103, 2017.

\bibitem{Cui:2019dwv}
Zhu-Fang Cui, Jin-Li Zhang, Daniele Binosi, Feliciano de~Soto, C\'edric Mezrag,
  Joannis Papavassiliou, Craig~D Roberts, Jose Rodr\'\i{}guez-Quintero, Jorge
  Segovia, and Savvas Zafeiropoulos.
\newblock {Effective charge from lattice QCD}.
\newblock {\em Chin. Phys. C}, 44(8):083102, 2020.

\bibitem{Gribov:1984tu}
L.V. Gribov, E.M. Levin, and M.G. Ryskin.
\newblock {Semihard Processes in QCD}.
\newblock {\em Phys. Rept.}, 100:1--150, 1983.

\bibitem{Mueller:1985wy}
Alfred~H. Mueller and Jian-wei Qiu.
\newblock {Gluon Recombination and Shadowing at Small Values of x}.
\newblock {\em Nucl. Phys. B}, 268:427--452, 1986.

\bibitem{Chen:2013nga}
Xurong Chen, Jianhong Ruan, Rong Wang, Pengming Zhang, and Wei Zhu.
\newblock {Applications of a nonlinear evolution equation I: the parton
  distributions in the proton}.
\newblock {\em Int. J. Mod. Phys. E}, 23(10):1450057, 2014.

\bibitem{Chang:2014lva}
Lei Chang, C\'edric Mezrag, Herv\'e Moutarde, Craig~D. Roberts, Jose
  Rodr\'\i{}guez-Quintero, and Peter~C. Tandy.
\newblock {Basic features of the pion valence-quark distribution function}.
\newblock {\em Phys. Lett. B}, 737:23--29, 2014.

\bibitem{Bourrely:2018yck}
Claude Bourrely and Jacques Soffer.
\newblock {Statistical approach of pion parton distributions from
  Drell\textendash{}Yan process}.
\newblock {\em Nucl. Phys. A}, 981:118--129, 2019.

\bibitem{Bourrely:2020izp}
Claude Bourrely, Franco Buccella, and Jen-Chieh Peng.
\newblock {A new extraction of pion parton distributions in the statistical
  model}.
\newblock {\em Phys. Lett. B}, 813:136021, 2021.

\bibitem{Wang:2014lua}
Rong Wang and Xurong Chen.
\newblock {Valence quark distributions of the proton from maximum entropy
  approach}.
\newblock {\em Phys. Rev. D}, 91:054026, 2015.

\bibitem{Han:2018wsw}
Chengdong Han, Hanyang Xing, Xiaopeng Wang, Qiang Fu, Rong Wang, and Xurong
  Chen.
\newblock {Pion Valence Quark Distributions from Maximum Entropy Method}.
\newblock {\em Phys. Lett. B}, 800:135066, 2020.

\bibitem{Dokshitzer:1977sg}
Yuri~L. Dokshitzer.
\newblock {Calculation of the Structure Functions for Deep Inelastic Scattering
  and e+ e- Annihilation by Perturbation Theory in Quantum Chromodynamics.}
\newblock {\em Sov. Phys. JETP}, 46:641--653, 1977.

\bibitem{Gribov:1972ri}
V.N. Gribov and L.N. Lipatov.
\newblock {Deep inelastic e p scattering in perturbation theory}.
\newblock {\em Sov. J. Nucl. Phys.}, 15:438--450, 1972.

\bibitem{Altarelli:1977zs}
Guido Altarelli and G.~Parisi.
\newblock {Asymptotic Freedom in Parton Language}.
\newblock {\em Nucl. Phys. B}, 126:298--318, 1977.

\bibitem{Zhu:1998hg}
Wei Zhu.
\newblock {A New approach to parton recombination in a QCD evolution equation}.
\newblock {\em Nucl. Phys. B}, 551:245--274, 1999.

\bibitem{Zhu:1999ht}
Wei Zhu and Jian-hong Ruan.
\newblock {A New modified Altarelli-Parisi evolution equation with parton
  recombination in proton}.
\newblock {\em Nucl. Phys. B}, 559:378--392, 1999.

\bibitem{Zhu:2004xj}
Wei Zhu and Zhen-qi Shen.
\newblock {Properties of the gluon recombination functions}.
\newblock {\em HEP. \& NP. Vol.}, 2:,109--114, 2005.

\bibitem{Pumplin:2001ct}
J.~Pumplin, D.~Stump, R.~Brock, D.~Casey, J.~Huston, J.~Kalk, H.~L. Lai, and
  W.~K. Tung.
\newblock {Uncertainties of predictions from parton distribution functions. 2.
  The Hessian method}.
\newblock {\em Phys. Rev. D}, 65:014013, 2001.

\bibitem{Martin:2002aw}
A.~D. Martin, R.~G. Roberts, W.~J. Stirling, and R.~S. Thorne.
\newblock {Uncertainties of predictions from parton distributions. 1:
  Experimental errors}.
\newblock {\em Eur. Phys. J. C}, 28:455--473, 2003.

\bibitem{Chen:2018wyz}
Xurong Chen.
\newblock {A Plan for Electron Ion Collider in China}.
\newblock {\em PoS}, DIS2018:170, 2018.

\bibitem{Accardi:2012qut}
A.~Accardi et~al.
\newblock {Electron Ion Collider: The Next QCD Frontier}: {Understanding the
  glue that binds us all}.
\newblock {\em Eur. Phys. J. A}, 52(9):268, 2016.

\bibitem{Brodsky:1973kr}
Stanley~J. Brodsky and Glennys~R. Farrar.
\newblock {Scaling Laws at Large Transverse Momentum}.
\newblock {\em Phys. Rev. Lett.}, 31:1153--1156, 1973.

\bibitem{Ball:2016spl}
Richard~D. Ball, Emanuele~R. Nocera, and Juan Rojo.
\newblock {The asymptotic behaviour of parton distributions at small and large
  $x$}.
\newblock {\em Eur. Phys. J. C}, 76(7):383, 2016.

\bibitem{Regge:1959mz}
T.~Regge.
\newblock {Introduction to complex orbital momenta}.
\newblock {\em Nuovo Cim.}, 14:951, 1959.

\bibitem{Best:1997qp}
C.~Best, M.~Gockeler, R.~Horsley, Ernst-Michael Ilgenfritz, H.~Perlt, Paul~E.L.
  Rakow, A.~Schafer, G.~Schierholz, A.~Schiller, and S.~Schramm.
\newblock {Pion and rho structure functions from lattice QCD}.
\newblock {\em Phys. Rev. D}, 56:2743--2754, 1997.

\bibitem{Lin:2020ssv}
Huey-Wen Lin, Jiunn-Wei Chen, Zhouyou Fan, Jian-Hui Zhang, and Rui Zhang.
\newblock {Valence-quark distribution of the kaon and pion from lattice QCD}.
\newblock {\em Phys. Rev. D}, 103(1):014516, 2021.

\bibitem{Davidson:1994uv}
R.M. Davidson and E.~Ruiz~Arriola.
\newblock {Structure functions of pseudoscalar mesons in the SU(3) NJL model}.
\newblock {\em Phys. Lett. B}, 348:163--169, 1995.

\bibitem{piIMParton-github}
{lukeronger/piIMParton}.
\newblock \url{https://github.com/lukeronger/piIMParton}.
\newblock Accessed: 2020-10-26.

\bibitem{Denisov:2018unj}
B.~Adams et~al.
\newblock {Letter of Intent: A New QCD facility at the M2 beam line of the CERN
  SPS (COMPASS++/AMBER)}.
\newblock 8 2018.

\bibitem{JLabPionProposal}
{Measurement of Tagged Deep Inelastic Scattering (TDIS)}.
\newblock \url{https://www.jlab.org/exp_prog/proposals/15/PR12-15-006.pdf}.
\newblock Accessed: 2020-10-26.

\end{thebibliography}

\end{document}